\documentclass[9pt,article,twoside]{rmaa-rho-class/rmaa-rho}
\RMxAAtemplatetype{\RMxAA} 

\vol{61}
\pages{30-36}
\thisyear{2025}
\doi{\href{https://www.astroscu.unam.mx/rmaa/RMxAA..XX-X}{https://www.astroscu.unam.mx/rmaa/RMxAA..XX-X}}

\title{AGN Variability with Rubin Observatory in the 2030s}

\author[,1]{S.~Panda \orcidlink{0000-0002-5854-7426}\thanks{Gemini Science Fellow}}
\author[2]{F.~Pozo Nu\~nez \orcidlink{0000-0002-6716-4179}}
\author[3]{H.~Benati Gon\c{c}alves \orcidlink{0009-0006-0492-9679}}
\author[4]{G.~Li \orcidlink{0000-0003-4007-5771}}
\author[5]{B.~Czerny \orcidlink{0000-0001-5848-4333}}
\author[6,7]{P.~Marziani \orcidlink{0000-0002-6058-4912}}
\author[3]{T.~Storchi-Bergmann \orcidlink{0000-0003-1772-0023}}

\affil[1]{International Gemini Observatory/NSF NOIRLab, Casilla 603, La Serena, Chile}
\affil[2]{Astroinformatics, Heidelberg Institute for Theoretical Studies, Schloss-Wolfsbrunnenweg 35, 69118 Heidelberg, Germany}
\affil[3]{Departamento de Astronomia, Instituto de Física, Universidade Federal do Rio Grande do Sul, CP 15051, 91501-970, Porto Alegre, RS, Brazil}
\affil[4]{Kavli Institute for Astronomy and Astrophysics, Peking University, Beijing 100871, People's Republic of China}
\affil[5]{Center for Theoretical Physics, Polish Academy of Sciences, Al. Lotnik\'ow 32/46, 02-668 Warsaw, Poland}
\affil[6]{Instituto de Astrofísica de Andalucía (IAA-CSIC), Glorieta de la Astronomía s/n, 18008 Granada, Spain}
\affil[7]{Istituto Nazionale di Astrofisica, Osservatorio di Padova, Vicolo dell'Osservatorio 5, 35122, Padova, Italy}

\leadauthor{S. Panda et al.}
\smalltitle{AGN Variability with LSST}

\corres{Swayamtrupta Panda}
\email{swayamtrupta.panda@noirlab.edu}

\received{January 30, 2026}
\accepted{\today}

\license{Texto de la licencia aquí}

\setbool{rho-abstract}{true} 
\setbool{rho-resumen}{true} 

\begin{abstract}
AGN variability offers a direct probe of accretion physics, disk structure, and black hole growth, but progress has been limited by sample size, cadence heterogeneity, and photometric systematics. The Vera C. Rubin Observatory Legacy Survey of Space and Time (LSST) will deliver multi-band light curves for millions of AGN, enabling variability studies at a true population scale. We synthesize recent results from the Zwicky Transient Facility (ZTF), which demonstrate that optical variability amplitudes and timescales are primarily regulated by accretion state, with secondary dependence on black hole mass and redshift, and establish the feasibility of survey-driven continuum reverberation mapping. ZTF measurements reveal optical continuum-emitting region sizes that often exceed standard thin-disk predictions, implicating diffuse continuum emission from the broad-line region as a significant contributor to observed inter-band lags. We evaluate the implications of LSST cadence and survey strategy—particularly deep-drilling configurations—for continuum and emission-line reverberation mapping, changing-look AGN, extreme variability quasars, and periodic variability searches. Key limitations of broadband photometric variability are identified, including variable emission-line contamination, diffuse BLR continuum emission, and cadence-dependent lag recoverability. We argue that realizing LSST’s full scientific potential requires community-scale, standardized variability metric pipelines, probabilistic classification integrated with alert brokers for follow-up triggering, and complementary medium-band photometric observations to isolate the accretion-disk continuum. Together, these elements will enable LSST to convert photometric variability into quantitative constraints on accretion disks, BLR structure, and supermassive black hole growth across cosmic time.
\end{abstract}


\begin{resumen}
La variabilidad de los núcleos galácticos activos (AGN) proporciona una vía directa para estudiar la física de la acreción, la estructura del disco y el crecimiento de los agujeros negros. Sin embargo, el progreso en este campo ha estado limitado por el tamaño de las muestras, la heterogeneidad en la cadencia observacional y las sistemáticas fotométricas. El Legacy Survey of Space and Time (LSST) del Observatorio Vera C. Rubin proporcionará curvas de luz multibanda para millones de AGN, permitiendo por primera vez estudios de variabilidad a escala poblacional. Sintetizamos resultados recientes del Zwicky Transient Facility (ZTF), que muestran que las amplitudes y escalas temporales de la variabilidad óptica están reguladas principalmente por el estado de acreción, con una dependencia secundaria de la masa del agujero negro y del corrimiento al rojo, y demuestran la viabilidad del mapeo de reverberación del continuo basado en programas fotométricos de gran campo. Las mediciones de ZTF revelan tamaños de las regiones emisoras del continuo óptico que a menudo superan las predicciones del modelo estándar de disco delgado, lo que sugiere que la emisión difusa del continuo procedente de la región de líneas anchas contribuye de forma significativa a los retardos observados entre bandas. Evaluamos las implicaciones de la cadencia y de la estrategia de observación de LSST, en particular en los campos profundos, para el mapeo de reverberación del continuo y de líneas de emisión, los AGN de tipo changing-look, los cuásares de variabilidad extrema y la búsqueda de variabilidad periódica. Identificamos limitaciones clave inherentes a la variabilidad fotométrica de banda ancha, incluyendo la contaminación variable por líneas de emisión, la contribución de la emisión difusa del continuo de la región de líneas anchas y la dependencia de la recuperación de retardos de la cadencia observacional. Argumentamos que explotar plenamente el potencial científico de LSST requiere métricas de variabilidad estandarizadas para toda la comunidad, clasificaciones probabilísticas integradas con alert brokers para la activación eficiente de seguimientos y observaciones fotométricas complementarias de banda media que permitan aislar el continuo del disco de acreción. En conjunto, estos elementos permitirán que LSST transforme la variabilidad fotométrica en restricciones cuantitativas sobre los discos de acreción, la estructura de la región de líneas anchas y el crecimiento de los agujeros negros supermasivos a lo largo de la historia cósmica.
\end{resumen}

\begin{document}

\maketitle
\pagestyle{fancy}\thispagestyle{firststyle}


\section{INTRODUCTION}

\RMxAAstart{A}ctive galactic nuclei (AGN) variability has long been recognized as one of the most direct, information-rich probes of accretion physics and black hole–galaxy coevolution, yet it remains one of the lesser unified observational frontiers in extragalactic astrophysics. Variability is not a secondary property of AGN—it is a defining one, manifesting across the electromagnetic spectrum and over timescales from minutes to decades \citep{Cackett_2021iSci...24j2557C, Blandford_1982ApJ...255..419B, Sugunuma_2006ApJ...639...46S, Bentz_2009ApJ...697..160B, Fausnaugh_2016ApJ...821...56F, Kaspi_2021ApJ...915..129K, Panda_2024ApJ...968L..16P}. Stochastic continuum fluctuations trace the size, temperature structure, and energy transport within accretion disks; correlated line and continuum variability encodes the geometry and kinematics of the broad-line region \citep[BLR,][]{SS_1973A&A....24..337S, Horne_2004PASP..116..465H, Czerny_2023A&A...675A.163C, Neustadt_2024ApJ...961..219N, Hagen_2024MNRAS.530.4850H}; and rare, extreme transitions challenge our most basic assumptions about the stability of accretion flows and obscuring structures \citep{Noda_2018MNRAS.480.3898N, Sniegowska_2020A&A...641A.167S, LaMassa_2015ApJ...800..144L, Stern_2018ApJ...864...27S, Graham_2020MNRAS.491.4925G, Panda_2024ApJS..272...13P, Jana_2025A&A...693A..35J, Guo_2025ApJ...995..139G}. Despite decades of progress, variability-driven constraints on AGN structure remain fragmented, limited by sample sizes, heterogeneous cadences, and systematic uncertainties in photometric and spectroscopic measurements. The imminent start of the Vera C. Rubin Observatory Legacy Survey of Space and Time \citep[LSST,][]{LSST_2019ApJ...873..111I} marks a decisive inflection point\footnote{the first public alerts are planned for February 4th 2026.}: wherein, AGN variability will be studied as a population-level phenomenon across cosmic time, rather than as a collection of bespoke case studies \citep{Panda_2019FrASS...6...75P, Kovacevic_2022ApJS..262...49K, Czerny_2023A&A...675A.163C, FPN_2024RNAAS...8...47P, Li_2025arXiv251208654L}.

Over the past decade, wide-field time-domain surveys such as the Zwicky Transient Facility (ZTF; \citealt{ZTF_2019PASP..131a8002B, ZTF_2019PASP..131g8001G}) have provided a clear preview of the coming era of variability-selected AGN science. ZTF has delivered major advances in continuum reverberation mapping, extreme-variability quasars (EVQs; \citealt{Graham_2020MNRAS.491.4925G, Ren_2022ApJ...925...50R}), and changing-look and changing-state AGN \citep{LaMassa_2015ApJ...800..144L, Runnoe_2016MNRAS.455.1691R, Stern_2018ApJ...864...27S, MacLeod_2019ApJ...874....8M, SanchezSaez2021AJ....162..206S, Temple_2023MNRAS.518.2938T, Ricci_2023NatAs...7.1282R, Zeltyn_2024ApJ...966...85Z, Panda_2024ApJS..272...13P, ShuWang_2024ApJ...966..128W, Guo_DESI_2025ApJS..278...28G}, while also revealing tidal disruption events \citep[TDEs,][]{Rees1988Natur.333..523R, vanVelzen2020SSRv..216..124V, Gezari2021ARA&A..59...21G}, newborn \citep{Arevalo2024A&A...683L...8A, Sanchez-Saez2024A&A...688A.157S, Hernandez_Garcia2025NatAs...9..895H} and flaring AGNs \citep{Trakhtenbrot2019NatAs...3..242T, Frederick2021ApJ...920...56F}, ambiguous nuclear transients \citep[ANTs,][]{Holoien2022ApJ...933..196H, Hinkle2024MNRAS.531.2603H, Wiseman2025MNRAS.537.2024W}, and candidate supermassive black hole (SMBH) binaries \citep[SMBHBs,][]{Ju2013ApJ...777...44J, DOrazio2023arXiv231016896D, Charisi2022MNRAS.510.5929C, Fatovic2025A&A...695A.208F, Marziani2025Univ...11...76M}. These transient phenomena are particularly powerful probes of black hole demographics - especially in identifying/charaterizing intermediate mass black holes (IMBHs). IMBHs can be identified either through low-mass AGNs \citep{Mezcua2017IJMPD..2630021M, Greene2020ARA&A..58..257G, Bernal2025A&A...694A.127B} or via transient illumination of otherwise dormant black holes. Among these, TDEs—stellar disruptions by black holes with $M_{\rm BH}\lesssim10^{8}M_\odot$—provide a uniquely sensitive window onto quiescent SMBHs and IMBHs \citep{Rees1988Natur.333..523R, vanVelzen2020SSRv..216..124V, Gezari2021ARA&A..59...21G}. Related classes, including flaring AGNs \citep{Trakhtenbrot2019NatAs...3..242T, Frederick2021ApJ...920...56F} and ANTs (possibly TDEs of intermediate- or high-mass stars, or TDEs occurring in AGNs; \citealt{Wiseman2025MNRAS.537.2024W}), further blur traditional boundaries between accretion-driven and disruption-driven variability.

At the same time, ZTF has exposed the limitations of piecemeal approaches to discovery and classification: the science now spans heterogeneous populations, overlapping phenomenology, and widely varying timescales, yet relies on fragmented, survey-specific workflows \citep[see e.g.,][]{Gupta2025MNRAS.542L.132G}. These lessons point to the need for a coordinated, community-scale framework—a comprehensive, well-characterized database of AGN variability detections with robust metrics and probabilistic classifications, tightly integrated with alert brokers and follow-up infrastructure \citep{LSST_2019ApJ...873..111I}. Such a framework is not merely logistical; it is essential for translating LSST’s unprecedented alert stream into physical insight and for enabling informed target selection for intensive monitoring, spectroscopy, and multi-wavelength campaigns.

\begin{figure*}[!htb]
    \centering
    \includegraphics[width=\linewidth]{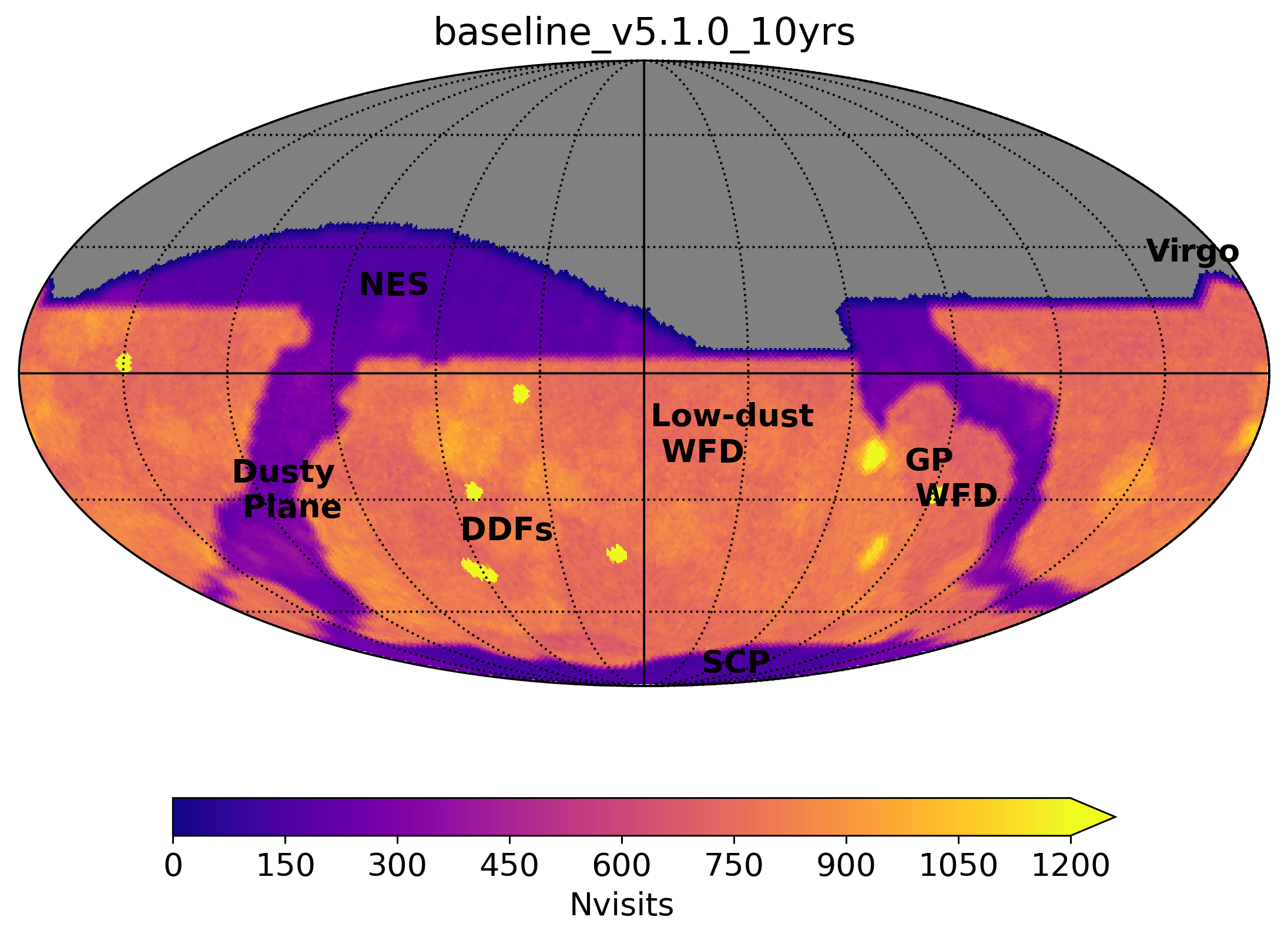}
    \includegraphics[width=\linewidth]{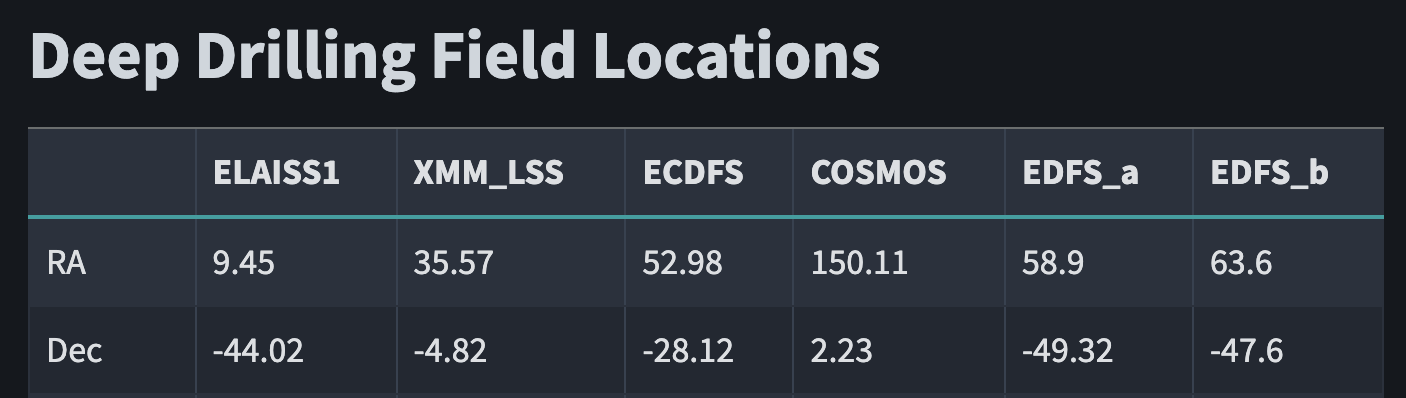}
    \caption{\justifying \textit{Top panel}: Survey footprint for the v5.1.0 cadence simulations. The auxiliary axis shows the number of planned visits over the 10-year baseline. \textit{Bottom panel}: The name and location of the six deep-drilling fields (DDFs) highlighted in the top panel.}
    \label{fig:footprint}
\end{figure*}

At the same time, the limitations of broadband photometry must be confronted head-on. Variable emission-line contamination, diffuse continuum emission from the BLR \citep{Lawther2018MNRAS.481..533L, Korista2019MNRAS.489.5284K, Chelouche2019NatAs...3..251C, FPN_2025A&A...700L...8P}, and cadence constraints all complicate the interpretation of LSST light curves as pure tracers of accretion disk variability \citep{FPN2023MNRAS.522.2002P, Czerny_2023A&A...675A.163C, FPN_2024RNAAS...8...47P}. These effects are not nuisances to be averaged over; they are integral to the physics of AGN and must be modeled explicitly if variability is to yield reliable structural constraints. Looking forward, a synergistic approach—combining LSST with targeted observations from meter-class telescopes equipped with carefully designed medium-band filters—offers a pragmatic path forward, balancing sensitivity, cadence, and spectral purity \citep[see e.g.,][]{Panda_2024ApJ...968L..16P}. In this short review, we frame AGN variability as a central pillar of LSST-era black hole astrophysics, outline the latest updates from the survey cadence simulations focusing on AGNs, key scientific opportunities and methodological challenges, and argue that the true legacy of Rubin Observatory will be defined not only by the volume of data it delivers, but by the coherence with which the community learns to interpret it.

\section{Rubin Observatory's LSST - ready to begin}

Rubin Observatory will elevate AGN variability studies from thousands to millions of sources, enabling systematic exploration across luminosity, black hole mass, accretion state, and redshift \citep{LSSTScienceBook2009arXiv0912.0201L, Brandt2018arXiv181106542B, LSST_2019ApJ...873..111I, Kovacevic_2022ApJS..262...49K, Czerny_2023A&A...675A.163C, FPN_2024RNAAS...8...47P}. At high redshift, LSST will probe variability-driven selection effects and apparent drop-offs in AGN populations, offering new leverage on duty cycles and growth histories \citep{Fan2006ARA&A..44..415F, Shankar2009ApJ...690...20S, Fan2023ARA&A..61..373F}. On short timescales, dense, multi-band light curves will allow rapid identification of sources suitable for high-cadence follow-up, opening a path to direct measurements of accretion disk sizes and temperature profiles via continuum reverberation mapping \citep{Fausnaugh_2016ApJ...821...56F, Homayouni2019ApJ...880..126H, Cackett_2021iSci...24j2557C, FPN2023MNRAS.522.2002P, Neustadt_2024ApJ...961..219N, Mandal2025arXiv251213296M}. These measurements will critically test the standard Shakura–Sunyaev disk paradigm, confront recent claims of disk size inflation or anomalous temperature gradients, and assess whether such discrepancies are universal, mass-dependent, or driven by unmodeled emission components. LSST will also bring intermediate-mass black holes into sharper focus, enabling scrutiny of the radius–luminosity relation in a previously inaccessible regime and motivating the adoption of novel, high-performance analysis techniques—such as GPU-accelerated and JAX-based inference frameworks—to handle the scale and complexity of the data \citep[see e.g.,][]{Yu2025arXiv251121479Y}.

Equally transformative will be LSST’s ability to extend variability baselines through careful inter-calibration with archival surveys and contemporaneous facilities \citep{Bon2016ApJS..225...29B, Li2025ApJ...988...42L, ElBadry2025arXiv250910601E}. Long-term light curves are indispensable for disentangling red noise from genuine secular evolution, for identifying periodic or quasi-periodic signals, and for characterizing rare but physically revealing phenomena such as changing-look transitions and extreme variability events \citep{Graham_2020MNRAS.491.4925G, Graham2023ApJ...942...99G, Komossa2024arXiv240800089K, Panda2025arXiv251001486P}. Achieving this goal, however, demands a rigorous understanding of photometric systematics, cross-survey calibration uncertainties, and band-dependent biases. Without such care, the very statistical power that defines LSST risks amplifying subtle systematics into dominant sources of error.

\subsection{LSST v5.1.0 Survey Strategy and Implications for AGN Variability}

The v5.1.0 LSST Operations Simulator (OpSim\footnote{\url{https://rubin-sim.lsst.io/}}) survey strategy represents a major step toward operational realism while largely preserving—and in many cases enhancing—the scientific ambitions of the Rubin Observatory Legacy Survey of Space and Time. Informed by early commissioning feedback, v5.1.0 adopts a single-snap visit model (1×30 s in \textit{grizy} and 1×38 s in \textit{u}), reflecting demonstrated LSSTCam\footnote{\url{https://lsstcam.lsst.io/}} performance and simplifying alert production, photometric calibration, and variability modeling. The simulations also incorporate more conservative slew and overhead times to better reflect expected on-sky efficiency during early operations. Despite these adjustments, v5.1.0 achieves slightly improved areal coverage, visit counts (see Figure \ref{fig:footprint}), and co-added depths relative to earlier baselines, with most science metrics showing net improvement.

\begin{figure}[!htb]
    \centering
    \includegraphics[width=\linewidth]{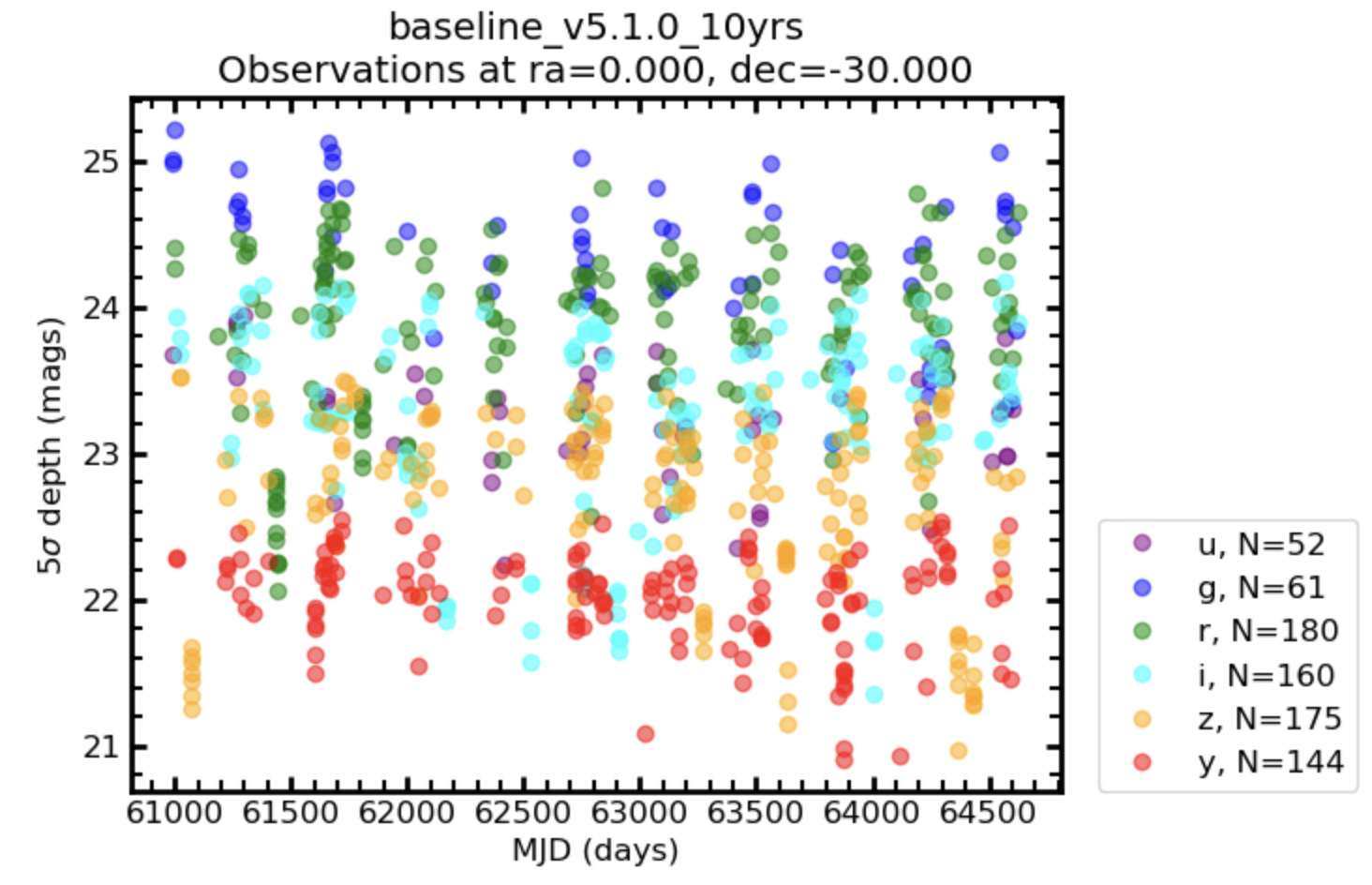}
    \caption{\justifying 5$\sigma$ depth as a function of time for a randomly selected location (RA = 0$^{\circ}$, Dec = -30$^{\circ}$) in the Wide-Fast-Deep (WFD) LSST footprint for the v5.1.0 baseline simulations over the 10 years. The six filters (\textit{ugrizy}) are shown using different colors. The legend highlights the number of visits corresponding to each filter.}
    \label{fig:wfd_lightcurve}
\end{figure}

\begin{figure}[!htb]
    \centering
    \includegraphics[width=\linewidth]{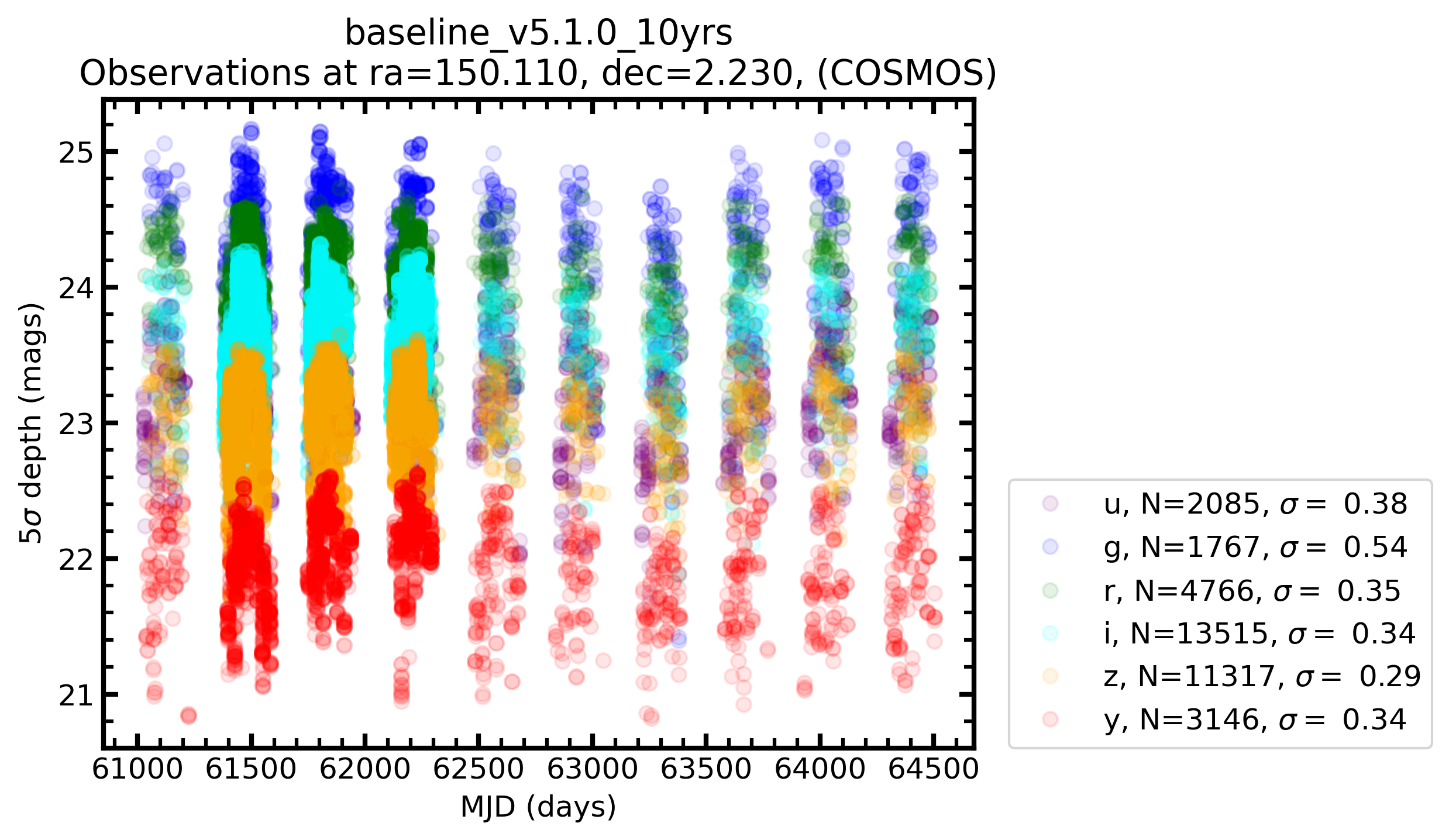}
    \caption{\justifying Similar to Figure \ref{fig:wfd_lightcurve}, but for COSMOS centered at RA = 150.11$^{\circ}$, Dec = 2.23$^{\circ}$ one of the six deep-drilling fields (DDFs). The legend highlights the number of visits and the standard deviation in the 5$\sigma$-depths corresponding to each filter. Notice the increase in the number of visits in the 2$^{\rm nd}$, 3$^{\rm rd}$, and 4$^{\rm th}$ year of operations in this field because of the adoption of the new ``ocean'' strategy.}
    \label{fig:ddf_lightcurve}
\end{figure}

\begin{figure}
    \centering
    \includegraphics[width=\linewidth]{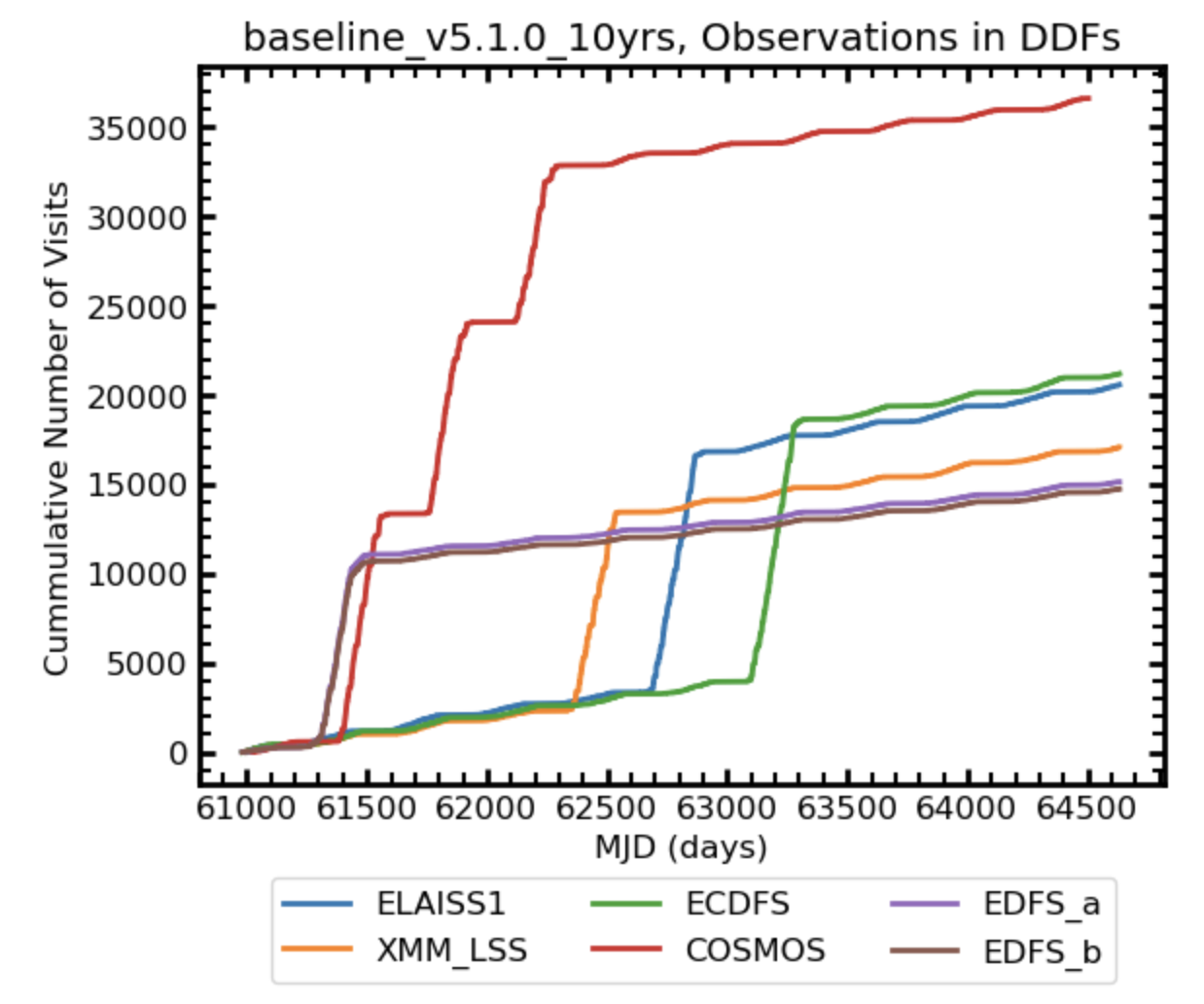}
    \caption{\justifying Cumulative number of visits in each deep-drilling field (DDF) based on the v5.1.0 baseline simulations over the 10 years. Notice the different season(s) of boosting in the number of visits for the DDFs. Careful optimization of science cases requiring dense cadence in each field is suggested.}
    \label{fig:cum_ddf}
\end{figure}

The most substantial evolution in cadence occurs in the Deep Drilling Fields (DDFs, see Figure \ref{fig:footprint}). Earlier strategies relied on long, sparsely spaced sequences that maximized depth but provided limited sensitivity to short-timescale variability. We demonstrate the stark difference in the increased number of visits in the DDFs relative to the at-large Wide-Fast-Deep (WFD) footprint in Figures \ref{fig:wfd_lightcurve} and \ref{fig:ddf_lightcurve}. In v5.1.0, the survey adopts the ``ocean” strategy, in which DDFs are observed frequently throughout most seasons using short sequences (typically $\sim$10 visits per night, every night or every other night), supplemented by at least one deep season per field featuring very long, high-cadence sequences ($\gtrsim$100 visits per filter). While the total number of DDF visits over the 10-year survey remains comparable to earlier plans, their redistribution in time dramatically enhances sensitivity to variability on timescales from days to months, while still preserving the ability to reach extreme depths (see Figure \ref{fig:cum_ddf}). A suite of cadence variants, including accordion modes, early deep seasons, rolling strategies, dithering experiments, and Year-1 template-focused runs, explores trade-offs between early science return, calibration robustness, and long-term uniformity\footnote{for more details on the LSST's observing strategy, the readers may refer to \url{https://survey-strategy.lsst.io/baseline/changes.html}}.

These design choices are particularly well aligned with the requirements of AGN variability science, which is explicitly recognized in v5.1.0 as a core survey driver. The high-frequency, multi-band sampling of the DDFs shall substantially improve the feasibility and precision of continuum reverberation mapping, enabling dense, season-long light curves for large samples of AGNs, while the deep seasons will provide the signal-to-noise needed for accurate lag measurements and accretion-disk size constraints. The improved temporal continuity will also enhance sensitivity to extreme-variability quasars and changing-look or changing-state AGN, allowing transitions to be detected earlier and followed more completely through their evolution. At longer timescales, the combination of frequent sampling and decade-long baselines will strengthen searches for periodic and quasi-periodic variability, narrowing the viable parameter space for supermassive black hole binary candidates while mitigating aliasing and red-noise confusion.

Taken together, the v5.1.0 strategy positions LSST to elevate AGN variability studies from heterogeneous, event-driven discoveries to a coherent, population-scale enterprise. Realizing this potential, however, will require parallel investment in classification pipelines, alert brokers\footnote{\url{https://rubinobservatory.org/for-scientists/data-products/alerts-and-brokers}}, and coordinated follow-up infrastructures capable of transforming LSST’s unprecedented time-domain data stream into physical insight.

\subsection{`Ocean' survey strategy in the DDFs}

While LSST will robustly populate the AGN parameter space across redshift, luminosity, and black hole mass, earlier works, e.g., \citet{FPN_2024RNAAS...8...47P}, demonstrated that the recoverability of reverberation signals is strongly cadence-dependent, particularly for continuum reverberation mapping (CRM). Using realistic baseline DDF cadences, they show that accretion-disk time delays can be reliably recovered primarily for massive systems ($\sim$10$^8$–10$^9$ M$_{\odot}$) at 1.5 $\lesssim$ z $\lesssim$ 2, where intrinsic disk lags are sufficiently long. For lower-mass black holes ($\lesssim$5 $\times$ 10$^7$ M$_{\odot}$) and lower redshifts, under-sampling leads to large uncertainties or failed lag recovery, especially in bands close to the driving continuum. Extending CRM to lower masses and luminosities requires near-daily ($\sim$1–2 nights) multi-band sampling over contiguous seasons, motivating dense, seasonally concentrated cadence designs.

In this context, the v5.1.0 ``ocean” strategy, specifically designed for the DDFs, represents a significant advance. By redistributing visits into frequent short sequences ($\sim$10 visits per night every night or every other night), punctuated by at least one deep, high-cadence season ($\leq$100 visits per filter), the ocean strategy substantially improves sensitivity to short- and intermediate-timescale variability while preserving total depth. For CRM, this directly mitigates the under-sampling limitations identified by Pozo Nuñez et al. and other works, increasing the fraction of quasars—particularly at intermediate masses and redshifts—for which disk lags may be measurable. For emission-line reverberation mapping, whose characteristic timescales are longer, the combination of frequent monitoring and deep seasons should enhance both lag detection and precision when paired with spectroscopic follow-up. Together with LSST’s immense and cadence-robust AGN census, the ocean strategy and its variants (accordion modes, early deep seasons, rolling cadence configurations) provide a realistic pathway toward scaling reverberation mapping into a population-level probe of AGN structure across cosmic time, discovery of variability-aided quasars, and bringing us closer to a complete census of quasars across redshifts.

\subsection{Overall quasar number counts}

Quasar number counts constitute one of the most direct empirical constraints on the demographics and cosmic evolution of accreting supermassive black holes. As functions of luminosity, redshift, and selection band, they encode the convolution of the black hole mass function, Eddington ratio distribution, and duty cycle, and thus provide the primary observational link between quasar phenomenology and black hole growth histories \citep{Hopkins2007ApJ...654..731H, Assef2011ApJ...728...56A, Fiore2017A&A...601A.143F, Shen2020MNRAS.495.3252S, Peca2024ApJ...974..156P}. Measurements of quasar number counts and luminosity functions have underpinned essentially all inferences about the rise and decline of luminous accretion over cosmic time, the apparent high-redshift turnover in quasar space density, and the efficiency with which black holes convert mass into radiation.

In the Rubin Observatory era, the diagnostic power of quasar number counts will be fundamentally transformed. LSST will deliver time-domain–selected samples of millions of quasars with well-characterized variability, extending well below the luminosity limits of spectroscopic surveys and to redshifts beyond those accessible to traditional color selection. Variability selection introduces a new and physically motivated axis in quasar demographics, preferentially sensitive to accretion state, stochastic disk fluctuations, and transient phases of black hole growth. As a result, LSST number counts will no longer reflect only the time-averaged luminosity function, but will probe the distribution of accretion episodes, duty cycles, and state transitions across the black hole population. Interpreting these counts will therefore require forward modeling that couples variability statistics to luminosity functions, enabling direct constraints on how often black holes turn on, how long they remain active, and how these processes evolve with mass and redshift.

\begin{figure*}[!htb]
    \centering
    \includegraphics[width=\linewidth]{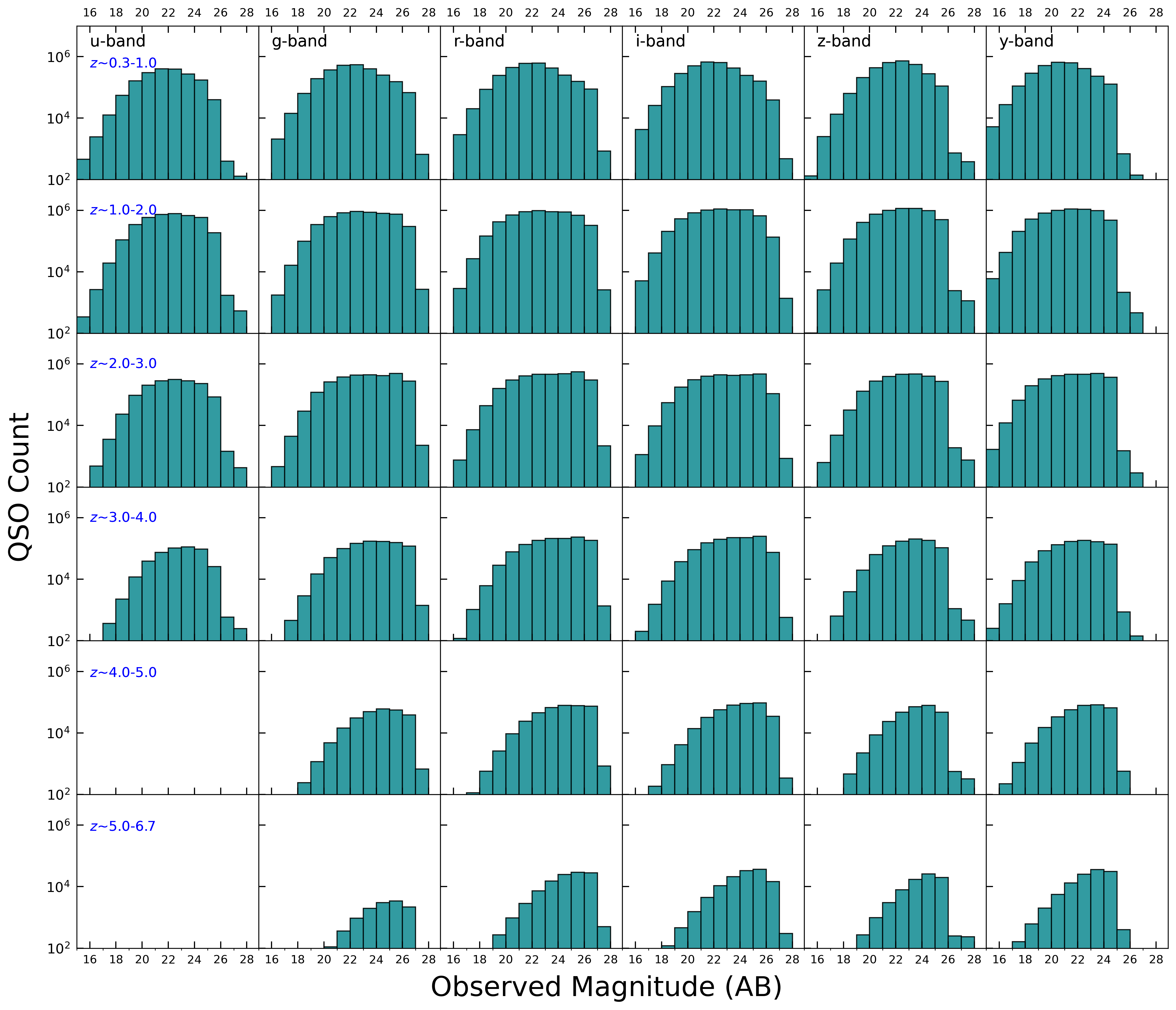}
    \caption{\justifying The distribution of QSO counts as a function of observed magnitude (AB) in different redshift ranges and filters, from the v5.1.0 simulations based on \citet{Li_2025arXiv251208654L}. Each column represents a specific filter ($u$, $g$, $r$, $i$, $z$, and $y$ from left to right), and each row corresponds to a redshift range. The blue bins indicate the number of QSOs detected in each magnitude bin. Adapted from Figure 5 in their paper.}
    \label{fig:qso_counts}
\end{figure*}

\citet{Li_2025arXiv251208654L} present a comprehensive forecast of the quasar and AGN populations detectable by LSST, using the v4.0 OpSim survey simulations and the Metrics Analysis Framework \citep[MAF,][]{Jones2014SPIE.9149E..0BJ} to quantify the impact of cadence choices on source counts. Adopting the \citet{Shen2020MNRAS.495.3252S} quasar luminosity function and a conservative luminosity cut (M$i$ $<$ -20), they predict that LSST will detect $\sim$6-12 million quasars across \textit{ugrizy} over 10 years, with the i-band yielding the largest sample, and that $\gtrsim$70\% of quasars will be detected within the first survey year as LSST rapidly reaches the break of the luminosity function over most redshifts. Extending beyond luminous quasars, they estimate $\sim$180-200 million lower-luminosity AGNs detectable in r and z bands based on deep-field number counts, with quasars comprising only $\sim$6\% of the total optically detected AGN population. Crucially, the study finds that most cadence variations—rolling strategies, DDF implementations, weather, and target-of-opportunity (ToO) scenarios—affect total quasar counts at the $\lesssim$1-2\% level, while u-band exposure-time variations yield only modest gains ($\lesssim$15\%), implying that LSST quasar demographics are remarkably robust to plausible survey-strategy changes, even as cadence details remain critical for time-domain and variability-driven AGN science. An updated version of their Figure 5 for the distribution of quasars as a function of observed magnitude in different redshift ranges and filters using the latest v5.1.0 simulations is shown in Figure \ref{fig:qso_counts}. As reported earlier, we find no difference in the overall QSO number counts between the previous (v4.0) and latest (v5.1.0) survey cadence simulations.

In the next section, we dive into the Zwicky Transient Facility (ZTF) and how it has helped set the foundation for the upcoming LSST and other large/long variability surveys in the context of AGN studies, with a slight focus on revealing the sizes of accretion disks in AGNs across redshifts.

\section{ZTF as a Precursor Facility for AGN Continuum Reverberation Mapping}

\begin{figure}[!htb]
    \centering
    \includegraphics[width=\linewidth]{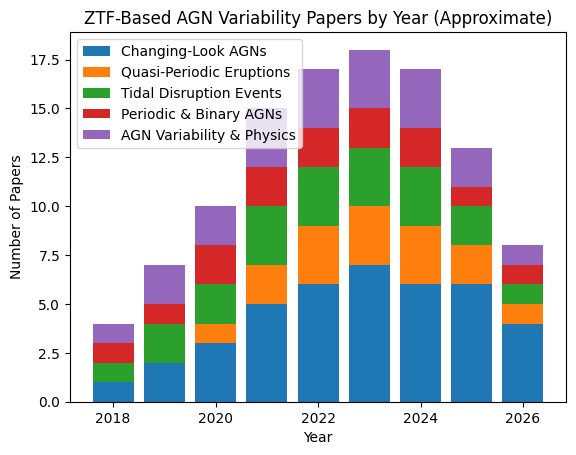}
    \caption{\justifying Stacked histogram showing the approximate number of ZTF-based AGN variability papers per year, broken down by five review categories from 2018–2026. Courtesy: \href{https://ui.adsabs.harvard.edu/}{NASA ADS service}.}
    \label{fig:ztf_papers}
\end{figure}

The Zwicky Transient Facility \citep[ZTF,][]{ZTF_2019PASP..131a8002B, ZTF_2019PASP..131g8001G} has established a new observational regime for optical time-domain studies of active galactic nuclei (AGN), providing a critical bridge between targeted reverberation-mapping campaigns and the forthcoming Vera C. Rubin Observatory Legacy Survey of Space and Time (LSST). In Figure \ref{fig:ztf_papers}, it can be clearly seen that the impact ZTF datasets have had in AGN variability-based studies across the 8+ years of its operations. With multiple broad-band (\textit{gri}) coverage, near-uniform cadences of a few days, and more than 7 years temporal baselines covering a footprint of the order of $\sim$27,500 square degrees across the northern hemisphere (about 3$\pi$ steradians of sky), ZTF has enabled the first statistically meaningful samples of AGN continuum reverberation measurements drawn from a survey-driven dataset \citep{Guo2022ApJ...929...19G, Guo2022ApJ...940...20G, Jha2022MNRAS.511.3005J, Su2025ApJ...990...10S}.

ZTF-based continuum reverberation mapping (CRM) has demonstrated that inter-band optical lags can be detected for hundreds of AGN, yielding empirical constraints on the characteristic sizes of the optical continuum-emitting region, although nuclear and host contamination can creep in with the use of broad-band datasets \citep[see Figure \ref{fig:ztf_cont} and Figure \ref{fig:ztf_cont_alt},][]{Panda_2024ApJ...968L..16P, FPN_2025A&A...700L...8P, Mandal2025arXiv251213296M}. These measurements consistently find continuum sizes that exceed the predictions of standard thin-disk models \citep{SS_1973A&A....24..337S}, corroborating earlier results from intensive single-object monitoring and microlensing studies to population scales \citep{Fausnaugh_2016ApJ...821...56F, Edelson2017ApJ...840...41E, Cackett2018ApJ...857...53C}. The size–luminosity relation inferred from ZTF CRM \citep{ShuWang2023ApJ...948L..23W} closely parallels the canonical H$\beta$ broad-line region (BLR) radius–luminosity relation \citep{Bentz_2009ApJ...697..160B, DallaBonta2020ApJ...903..112D}, supporting interpretations in which diffuse continuum emission (DCE) from the BLR contributes significantly to the observed optical lags \citep{Korista2001ApJ...553..695K, Lawther2018MNRAS.481..533L, Korista2019MNRAS.489.5284K, Chelouche2019NatAs...3..251C, Netzer2022MNRAS.509.2637N}. However, alternate scenarios have been discovered where, through the use of medium-band filters \citep{Panda_2024ApJ...968L..16P, FPN_2025A&A...700L...8P, Mandal2025arXiv251213296M} have allowed to recover the ``true'' accretion disk-emission, agreeing with the standard disk predictions, thus mitigating the contamination from the DCE and emission lines.

This empirical connection between continuum-emitting region size and BLR scale has motivated the use of CRM as a complementary pathway to BLR characterization and black hole mass estimation, particularly since spectroscopic reverberation mapping can be observationally expensive and limited to a few sources until large synoptic surveys can repeatedly spectroscopically monitor large areas of the sky. Findings from \citet{ShuWang_2024ApJ...966..128W} have shown that CRM-derived continuum sizes correlate tightly with BLR radii, enabling black hole mass estimates that are broadly consistent with traditional reverberation-based and single-epoch methods. In this framework, ZTF functions as a proof-of-concept for photometric reverberation methodologies that are expected to become central to LSST-era AGN studies.

At the same time, ZTF CRM analyses have highlighted the critical role of photometric fidelity in interpreting variability-based size measurements. Host-galaxy contamination, epoch-dependent PSF variations, reference-image systematics, and field-to-field calibration offsets can introduce spurious variability or suppress intrinsic signals, particularly for low-luminosity and low-redshift AGN. Studies employing custom photometric pipelines based on reference-subtracted images demonstrate that uncorrected systematics at the $\geq$0.05–0.1 mag level can significantly bias variability amplitudes, lag measurements, and inferred scaling relations \citep{Arevalo2025arXiv251006898A}\footnote{see also \url{http://nesssi.cacr.caltech.edu/ZTF/Web/Zuber.html}}. Accurate calibration, error modeling, and host subtraction are therefore essential prerequisites for extracting physically meaningful CRM constraints. 

In this sense, ZTF has played a dual role: it has delivered the first large-scale empirical constraints on AGN continuum reverberation from a wide-field time-domain survey, and it has exposed the calibration and analysis requirements necessary for scaling these methods to LSST. The methodologies developed and validated using ZTF—particularly those addressing photometric systematics—form a critical foundation for exploiting LSST to map accretion disk structure, BLR geometry, and black hole scaling relations across unprecedented AGN samples.

\subsection{Lessons learned?}

ZTF results demonstrate that the limiting factor for LSST AGN variability science will not be sensitivity or source counts, but the ability to systematically characterize, classify, and act on variability at scale. LSST will deliver well-sampled light curves for millions of AGN, making variability a primary observable for accretion physics, reverberation mapping, and transient accretion-state changes \citep[see][for a recent review]{Panda2025arXiv251001486P}. Extracting physical information from this data volume requires standardized, survey-wide variability metrics rather than object-specific analyses.

A community-maintained database of variability features, e.g., structure-function parameters, power spectral density (PSD) slopes and break frequencies, color-dependent variability, and lag proxies, is therefore essential. Such a database enables consistent population studies, cross-survey inter-calibration, and physically motivated source selection. ZTF experience shows that without unified metric pipelines, variability results become cadence-dependent, non-reproducible, and difficult to compare across samples.

This infrastructure is critical for follow-up-limited science. Continuum reverberation mapping, changing-look AGN, extreme variability quasars, and candidate SMBH binaries require early identification and prioritization while variability events are ongoing. LSST alert brokers\footnote{\url{https://rubinobservatory.org/for-scientists/data-products/alerts-and-brokers}} must therefore incorporate AGN-specific variability features and probabilistic classifications, enabling triggers based on inferred accretion state, expected lag timescales, or departures from historical variability behavior, rather than simple flux thresholds.

ZTF provides the empirical basis for this framework: it shows that variability-based AGN science scales only when detection, feature extraction, classification, and broker-driven triggering are tightly integrated. For LSST, this integration is not optional but foundational. Without it, LSST variability science will be dominated by systematics and follow-up inefficiencies; with it, LSST can convert its alert stream into a physically interpretable, population-level probe of AGN accretion and evolution.

\section{Accretion-State–Driven Variability and Reverberation Science with LSST}

The start of Rubin Observatory’s Legacy Survey of Space and Time (LSST), together with the adoption of the v5.1.0 ``ocean'' strategy for the Deep Drilling Fields (DDFs), establishes variability as a primary observable for AGN physics rather than a secondary byproduct of imaging depth. In this context, the complementary works of \citet{Hygor_2025ApJ...988...27B} and \citet{Panda2025arXiv251001486P} provide a physically motivated framework for interpreting LSST AGN variability in terms of accretion state, while clarifying the role of cadence in unlocking reverberation science at scale.

\begin{figure}[!htb]
    \centering
    \includegraphics[width=\linewidth]{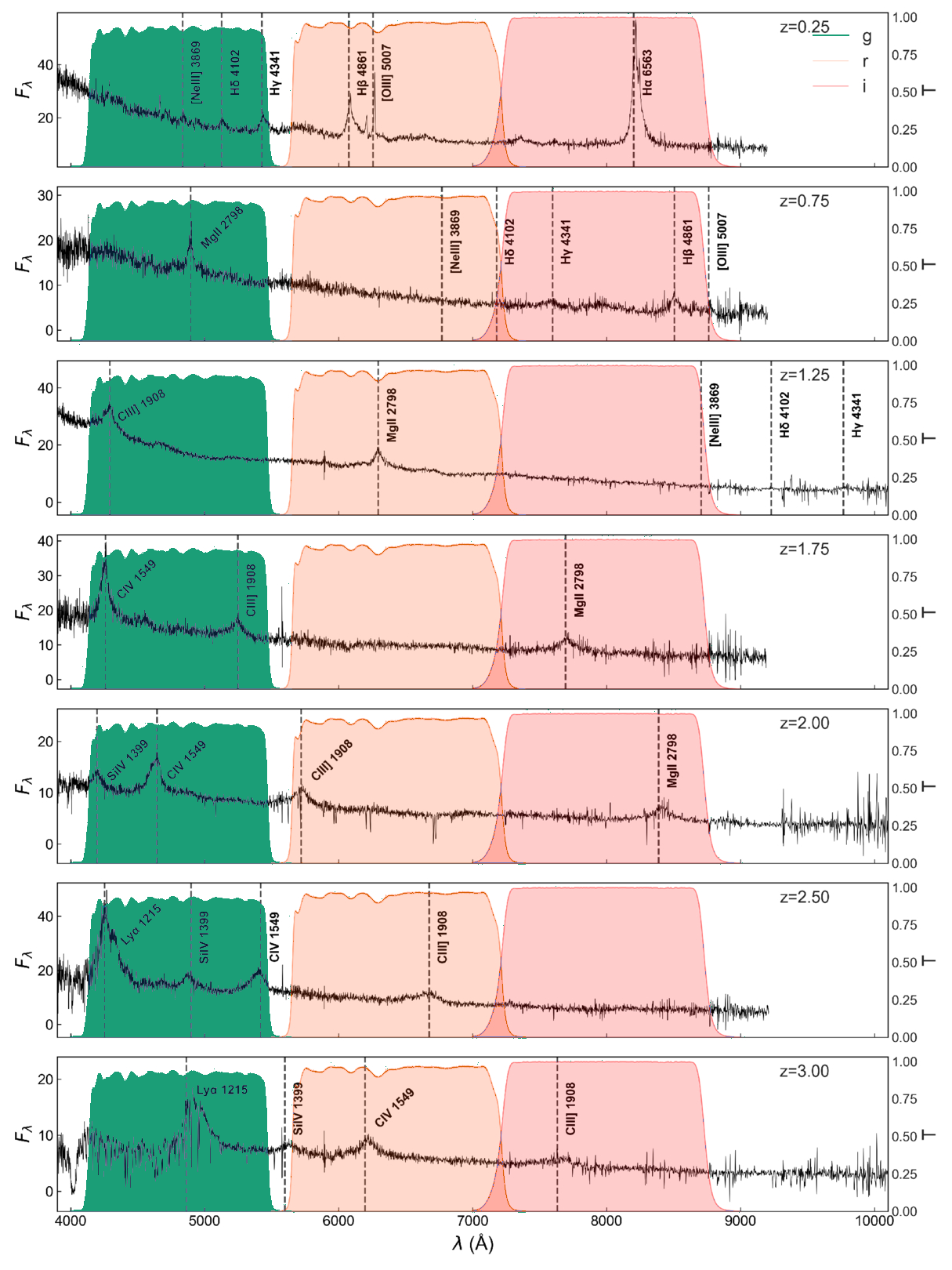}
    \caption{\justifying Representative SDSS spectra from our sample are shown for seven redshift bins: 0–0.25, 0.25–1.0, 0.75–1.25, 1.25–1.75, 1.75–2.0, 2.0–2.50, and 2.5–3.0. The ZTF filter transmission curves are overplotted, and prominent emission lines are labeled. Flux densities are given in units of 10$^{-17}$ erg s$^{-1}$ cm$^{-2}$ \AA$^{-1}$, while the filter transmission fraction (T) is shown on the right-hand y-axis. Courtesy: \citet{Hygor_2025ApJ...988...27B}.}
    \label{fig:ztf_cont}
\end{figure}

In a recent review, \citet{Panda2025arXiv251001486P} argue that the Eddington ratio is the principal organizing parameter of AGN phenomenology, regulating continuum variability, spectral properties, and the structure and size of the accretion disk (their Section 2 and their Figure 1). Through standard thin-disk scalings such as the disk size–luminosity relation, this framework implies that both continuum and emission-line reverberation are governed primarily by accretion state rather than black hole mass alone—a perspective that is especially powerful for LSST-era surveys, where photometric variability amplitudes and timescales serve as indirect probes of disk stability and radiative efficiency. Strong empirical support for this picture is provided by \citet{Hygor_2025ApJ...988...27B}, who use six years of ZTF monitoring of 915 quasars to demonstrate a tight, nearly redshift-independent anti-correlation between optical variability amplitude and Eddington ratio (their Figs. 3 and 5). Once the Eddington ratio is included, residual dependencies on black hole mass and redshift become weak, showing that accretion state dominates variability behavior across cosmic time and establishing photometric variability as a scalable estimator of Eddington ratio that naturally explains the prevalence of extreme variability quasars and changing-look AGN in low-Eddington regimes (cf. their Section 4.3).

Placed in the context of LSST quasar demographics—where $\approx$6–12 million quasars are expected, and total yields are largely insensitive to survey-strategy variations at the 1–2\% level—these results imply that cadence choices do not determine whether LSST finds quasars, but instead control what physical information can be extracted from their variability. The ensemble trends identified by \citet{Hygor_2025ApJ...988...27B} therefore motivate a complementary analysis in the frequency domain, where variability can be directly tied to characteristic physical timescales. Power spectral density (PSD) methods provide this link by decomposing optical light curves into scale-dependent components and enabling quantitative constraints on break frequencies, slopes, and normalizations as functions of intrinsic source parameters, effectively translating population-level correlations into physically interpretable timescale scalings. PSD results further reinforce that quasar variability is not governed by a single characteristic timescale, but instead reflects a continuum of processes whose relative contributions depend on black hole mass, accretion state, and rest-frame wavelength—fully consistent with the statistical trends reported by \citet{Hygor_2025ApJ...988...27B}, while extending them by isolating the specific frequency regimes in which these dependencies emerge. Such analyses, however, require long, well-sampled light curves with controlled cadence and photometric systematics, a requirement first met at population scale by the Zwicky Transient Facility \citep[ZTF;][]{ZTF_2019PASP..131a8002B, ZTF_2019PASP..131g8001G}, which bridges ensemble variability studies and frequency-domain modeling and provides the direct observational and methodological precursor to LSST variability science.

In this context, the ocean strategy—by redistributing visits into frequent short sequences across most seasons and including at least one ultra-deep, high-cadence season per field—directly addresses the cadence limitations identified by reverberation and disk-size studies (e.g., \citealt{FPN_2024RNAAS...8...47P, Panda2025arXiv251001486P}; see Section 3). Dense, near-daily multi-band sampling substantially improves sensitivity to short and intermediate timescale variability, thereby expanding the region of parameter space over which continuum reverberation lags and accretion-disk size measurements can be recovered beyond only the most massive and luminous systems.

\subsection{QSO selection in photo-variability surveys - a case study}

Another equally important aspect is the pre-selection and scrutiny of AGNs across redshift for these temporal studies. This is crucial also to envision the choices of filters to ingest for further analyses in both continuum- and line-based photometric reverberation mapping. In our previous work \citep{Czerny_2023A&A...675A.163C}, we defined the redshift-dependent quasar selection for photometric reverberation mapping by explicitly accounting for the LSST filter transmission curves and their effective throughput. We focused on the \textit{g}, \textit{r}, \textit{i}, and \textit{z} bands and restricted the analysis to prominent broad emission lines: H$\beta$, Mg{\sc ii}, and C{\sc iv}, which have well-established radius–luminosity relations \citep{Bentz_2009ApJ...697..160B, Kaspi2021ApJ...915..129K, Prince2023A&A...678A.189P, Shen2024ApJS..272...26S}. To ensure robust line–continuum separation, we required that the redshifted line fall within the flat, high-throughput region of a given LSST bandpass, avoiding both the steep bandpass roll-off and the overlap regions between adjacent filters. This requirement translated into acceptable observed-frame wavelength ranges of 4100–5300 \AA\ (\textit{g}), 5700–6700 \AA\ nm (\textit{r}), 7100–8000 \AA\ (\textit{i}), and 8300–9100 \AA\ (\textit{z}), with configurations near band edges or overlap regions excluded. 

\begin{figure}
    \centering
    \includegraphics[width=\linewidth, trim={0 4cm 0 1cm},clip]{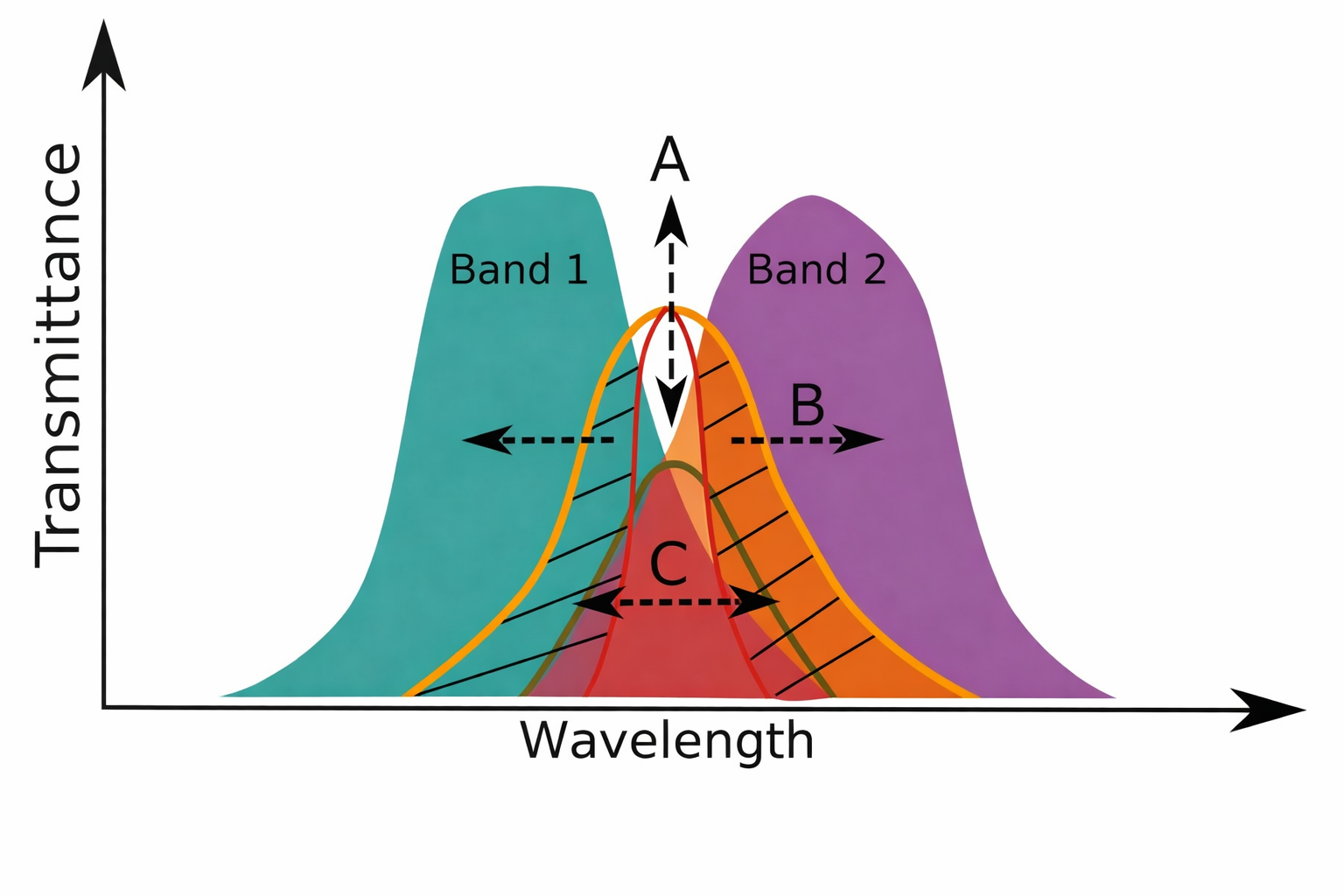}
    \caption{\justifying A representative illustration for the quasar selection based on distribution of (A) line intensities; (B) band overlaps, and (C) line widths. Three instances of emission line profiles are shown in orange, red, and green, and the patched region highlights the importance of the broad distribution of equivalent widths (EWs) in quasars. The arrows indicate the direction of the effect resulting from these three factors. Courtesy: \citet{Panda_2019FrASS...6...75P}.}
    \label{fig:panda19}
\end{figure}

At each redshift, if a redshifted emission line satisfied these throughput-based criteria, the corresponding filter was treated as the line-contaminated band, while a neighboring filter with negligible line contribution—selected based on its transmission curve—served as the uncontaminated continuum reference band; this procedure was automated. Modeling shows that partial truncation of the line profile by finite bandpasses introduces only small biases in the recovered delays ($<$5\% for symmetric clipping and $<$1\% for asymmetric coverage; \citealt{FPN2014A&A...568A..36P}). The mapping between emission lines and suitable LSST filters is inherently redshift dependent, and in some intervals multiple band–line combinations are possible (e.g., near z $\approx$ 0.5). Although inclusion of the \textit{z} band extends the analysis to higher redshifts, quasars in this regime are increasingly affected by broad absorption line systems and, at still higher redshifts, by Ly$\alpha$ forest absorption. We therefore limited our standard analysis to z $\leq$ 3.5, noting that for certain redshift ranges no suitable line satisfies the bandpass and throughput criteria because it falls too close to a filter edge or within an overlap region (see also \citealt{Panda_2019FrASS...6...75P} and Figure \ref{fig:panda19} for an adapted illustration). 

Looking ahead, the scientific return outlined above will depend critically on how LSST variability information is operationalized in real time. The combination of cadence-robust AGN demographics and accretion-state–sensitive variability metrics motivates the development of AGN-aware alert brokers that incorporate variability amplitude, structure-function evolution, and color-dependent variability as first-order features, rather than treating AGNs as background contaminants. Under the ocean DDF strategy, the availability of dense, multi-band light curves on nightly to near-nightly timescales enables early identification of accretion-state changes, allowing extreme variability quasars, changing-look AGN, and high-priority reverberation mapping candidates to be flagged while transitions are still in progress. Follow-up prioritization can then be guided by physically motivated criteria—such as inferred Eddington ratio, expected lag timescales, and seasonal visibility—maximizing the efficiency of spectroscopic and multi-wavelength campaigns. In this framework, the DDFs function not only as deep monitoring fields but also as training and validation sets for variability-based classifiers that can be exported to the WFD main survey, ensuring that LSST’s unprecedented alert stream is translated into a coherent, physics-driven AGN time-domain program.

\begin{figure}[!htb]
    \centering
    \includegraphics[width=\linewidth]{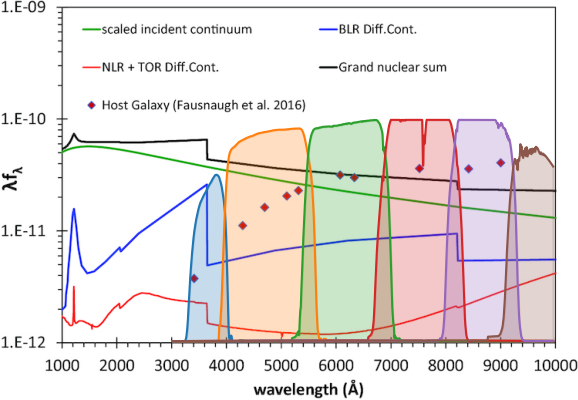}\\
    \includegraphics[width=\linewidth, trim={0 4.04cm 0 0},clip]{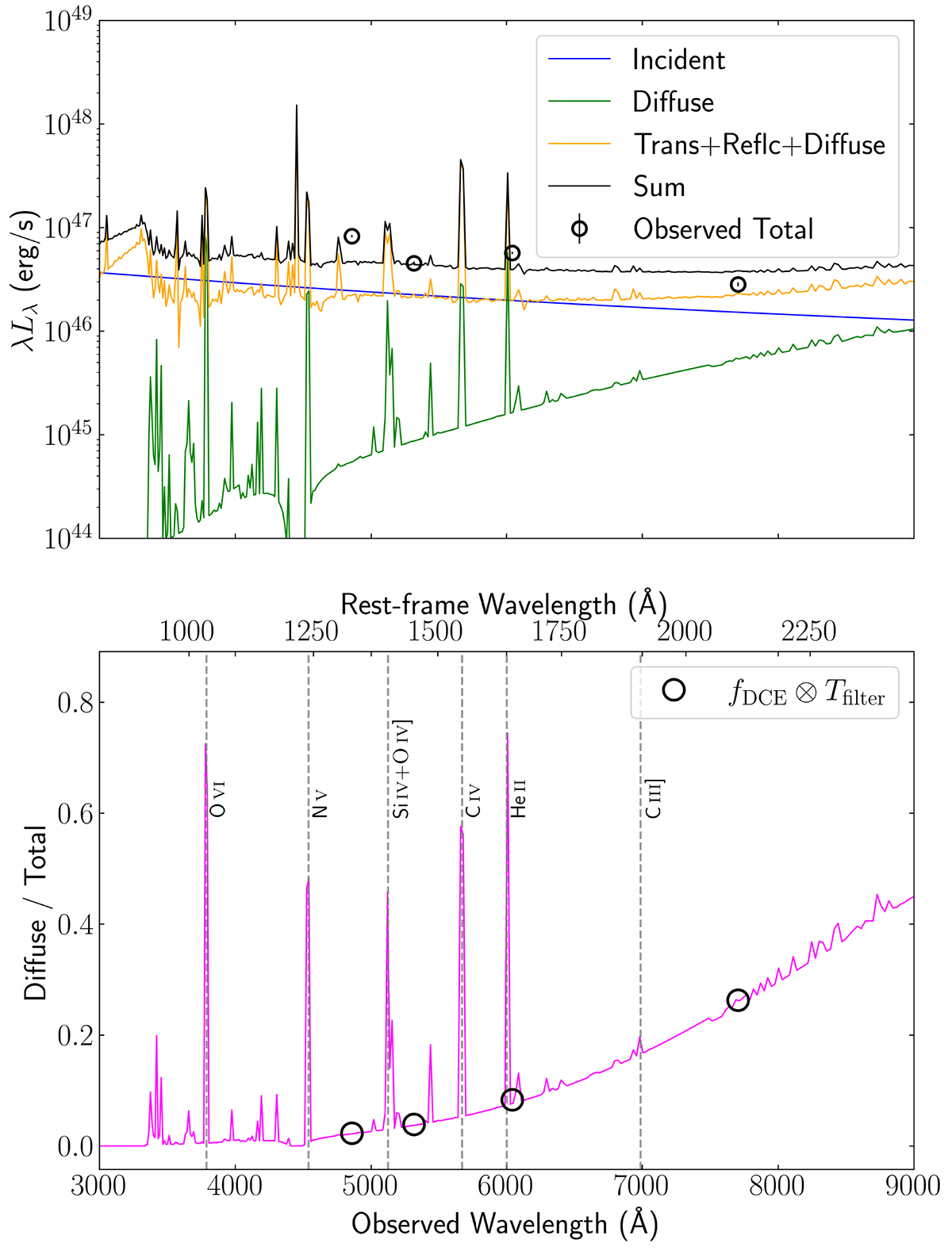}
    \caption{\justifying \textit{(Top:)} The computed diffuse-continuum (DC) spectra, in units of erg s$^{-1}$ cm$^{-2}$, from the inner and outer narrow-line regions of NGC~5548 and from dusty gas located near the graphite-grain sublimation radius (red) are shown, together with the BLR DC model from \citet{Korista2001ApJ...553..695K} (blue), an estimate of the underlying accretion-disc spectrum \citep{Magdziarz1998MNRAS.301..179M}, and their sum (black). The LSST 6-channel (\textit{ugrizy}) transmission throughput curves overlaid for reference to highlight the contribution of each underlying component. Courtesy: \citet{Korista2019MNRAS.489.5284K}. \textit{(Bottom:)} Top: The wavelength-dependent luminosity contributions from the various components are shown. The blue curve denotes the incident emission, corresponding to the intrinsic continuum irradiating the BLR clouds. The green curve shows the diffuse emission from photoionized gas within the BLR, while the orange curve represents the sum of the transmitted, reflected, and diffuse components. The black curve shows the total emission model (sum of all components), which is in good qualitative agreement with the observed total emission (black circles) derived from the photometric light curves. Courtesy: \citet{FPN_2025A&A...700L...8P}.}
    \label{fig:dce}
\end{figure}

\section{Looking Ahead}

The Rubin Observatory LSST will provide large, homogeneous samples of variable AGNs with long temporal baselines and multi-band photometric coverage, enabling a wide range of time-domain studies. At the same time, the cadence, wavelength coverage, and broadband nature of the data impose limitations that must be carefully considered when interpreting variability and reverberation signals. In this closing section, we summarize key challenges associated with broadband photometric variability, discuss observational and methodological approaches that can mitigate these effects, and outline infrastructure and data-access capabilities that will be important for precision AGN variability studies in the LSST era.

\subsection{Alert brokers and community-driven science}
The Rubin Observatory LSST will generate $\sim$10 million alerts per night, necessitating community alert brokers to enable real-time filtering, classification, and prioritization of variable and transient sources. Several community-developed brokers\footnote{for a full list of ``full-stream'' alert brokers and links to their websites/portals see \url{https://rubinobservatory.org/for-scientists/data-products/alerts-and-brokers}} are particularly well aligned with AGN variability, long-timescale monitoring, and photometric AGN selection (see Table \ref{tab:lsst_brokers_agn_compact}). Collectively, these brokers form a critical interface between LSST alert production and AGN science by enabling the community to (i) identify AGNs using variability and color information alone, (ii) distinguish stochastic AGN variability from transient phenomena, (iii) select rare subclasses such as changing-look AGNs and extreme variability quasars, and (iv) construct well-defined samples for reverberation mapping and line-contamination studies in the Deep Drilling Fields. Although many brokers share overlapping technical frameworks, each emphasizes distinct science priorities, underscoring the importance of transparent characterization of their capabilities for effective community adoption.

\subsection{Enabling data products for precision variability studies}
\begin{itemize}
\item \textit{Multi-wavelength SEDs and photometric redshifts:} Dedicated multi-wavelength SED analyses in several Rubin LSST Deep Drilling Fields (e.g., WCDFS, ELAIS-S1, XMM-LSS, and COSMOS) provide a critical foundation for future AGN variability, reverberation-mapping, and line-contamination studies. By leveraging homogeneous forced photometry from X-ray to far-infrared wavelengths and employing SED models that explicitly account for AGN emission and complex star-formation histories, these efforts enable robust characterization of host-galaxy properties and continuum components. Such SED baselines are essential for interpreting LSST light curves and improving photometric redshift accuracy \citep[see e.g.,][]{Zou2022ApJS..262...15Z}.

\item \textit{Need for calibrated image cutouts:} While LSST pipeline photometry will be sufficient for many applications, it may not deliver the precision required for studies of subtle effects such as low-amplitude variability or time-delayed line-contamination signals. Access to calibrated image cutouts (“postage stamps”) would enable custom photometric extraction tailored to individual sources without requiring the download of full multi-gigabyte images. Within the Rubin Science Platform, this functionality is described as an \emph{image cutout service}, often implemented via Virtual Observatory standards such as the IVOA Simple Observation Data Access (SODA) protocol. In this paradigm, users specify a region of interest, and the cutout is generated server-side and delivered directly to the client, minimizing unnecessary data transfer. A concrete demonstration is provided in the Rubin DP1 tutorial “103.4. Small image cutouts,” where users identify a full image via a datalink URL and retrieve only the requested subimage. Such cutout-level access to calibrated data has been discussed extensively within the Rubin Observatory community and is under active development.

\end{itemize}

\subsection{Efficient multi-wavelength follow-up strategies}
\begin{itemize}
\item The scientific return of LSST AGN variability studies will be further amplified through coordinated follow-up with ground- and space-based facilities spanning the radio to gamma-ray regimes, including SKA, JWST \citep{JWST_2006SSRv..123..485G}, CTA, Roman, 4MOST \citep{4MOST_2025ApJ...992..158F}, Euclid, SPHEREx, the Wide-field Spectroscopic Telescope (WST; \citealt{WST_2024arXiv240305398M}), and next-generation spectroscopic surveys such as Spec-S5 \citep{specs5_2025arXiv250307923B}. Strategic coordination across these facilities will enable comprehensive characterization of AGN accretion, environment, and feedback on timescales inaccessible to any single survey.
\end{itemize}

\subsection{Challenges of broadband photometric variability}
\begin{itemize}
\item \textit{Variable emission-line contamination:} Broadband fluxes include contributions from strong, time-variable emission lines whose strengths depend on redshift and filter bandpass. Single-epoch spectroscopy demonstrates that line contamination can significantly bias variability amplitudes and inter-band lag measurements when prominent lines enter or exit a given filter (see Figures \ref{fig:ztf_cont} and \ref{fig:panda19}), complicating the interpretation of broadband light curves as pure accretion-disk tracers.

\item \textit{Diffuse continuum emission from the BLR:} Reprocessed continuum emission originating in the broad-line region introduces additional lag components that are not associated with the intrinsic accretion-disk response. This diffuse continuum emission (DCE) can inflate inferred disk sizes and distort disk size--luminosity relations if not explicitly modeled \citep[see Figure \ref{fig:dce} top-panel,][]{Korista2019MNRAS.489.5284K, Chelouche2019NatAs...3..251C, Netzer2022MNRAS.509.2637N, FPN_2025A&A...700L...8P}. Because the DCE is a physical emission component rather than a removable systematic, and itself varies temporally, accurate interpretation requires source-by-source modeling informed by the intrinsic SED and the physical properties of the emitting medium, as illustrated in the bottom panel of Figure \ref{fig:dce}.

\item \textit{Cadence limitations for continuum reverberation mapping:} The recoverability of continuum lags depends sensitively on the sampling of characteristic disk timescales, which vary with black hole mass, accretion rate, and redshift. Simulations show that standard wide-field cadences undersample disk lags for large fractions of the AGN population, particularly at lower masses and redshifts, leading to biased or non-detections \citep{FPN2023MNRAS.522.2002P, FPN_2024RNAAS...8...47P}. Population-level lag recovery is therefore strongly coupled to survey cadence and source demographics \citep{Li_2025arXiv251208654L}, with boosted observing seasons—such as those enabled by the ocean strategy—playing a critical role.

\item \textit{Complementing broadband surveys with medium-band follow-up:} Targeted observations with meter-class telescopes equipped with custom medium-band filters provide a practical avenue for isolating accretion-disk continuum emission while minimizing contamination from emission lines and diffuse continuum emission \citep{Chelouche2019NatAs...3..251C, Panda_2024ApJ...968L..16P, Jaiswal2025A&A...702A..92J, FPN_2025A&A...700L...8P, Mandal2025arXiv251213296M}. By balancing spectral purity against exposure-time requirements, such follow-up enables recovery of physically meaningful continuum lags in regimes where broadband photometry alone yields biased disk-size estimates, and offers a scalable and cost-effective complement to LSST-era variability studies.

\end{itemize}

Rubin Observatory LSST will transform AGN variability into a precision tool for probing accretion physics by enabling large-scale, multi-band reverberation and variability studies across a broad parameter space. Fully realizing this potential requires explicitly modeling emission-line and diffuse BLR continuum contamination rather than treating them as systematics. A two-tier strategy -- LSST broadband monitoring combined with targeted medium-band follow-up—offers a robust path to unbiased disk size measurements and falsifiable tests of accretion-disk structure and evolution, and other diverse aspects of AGN variability.

\begin{table}
\centering
\caption{Rubin LSST alert brokers relevant to AGN variability and selection studies.}
\label{tab:lsst_brokers_agn_compact}
\resizebox{\columnwidth}{!}{
\begin{tabular}{lcccc}
\hline
Broker &
AGN selection &
Long-term variability &
Extreme AGN detection &
Photo-RM \\
\hline
ANTARES & \checkmark & \checkmark & $\circ$ & \checkmark \\
ALeRCE  & \checkmark & \checkmark & \checkmark & $\circ$ \\
Fink    & $\circ$    & $\circ$    & \checkmark & $\circ$ \\
Lasair  & \checkmark & \checkmark & $\circ$ & \checkmark \\
AMPEL   & \checkmark & \checkmark & \checkmark & \checkmark \\
\hline
\end{tabular}
}
\tablecomments{$\checkmark$ = strong capability; $\circ$ = partial or use-case dependent.}
\end{table}

\renewcommand{\refname}{REFERENCES}
\bibliography{rmaa}

\begin{thebibliography}{}
\expandafter\ifx\csname natexlab\endcsname\relax\def\natexlab#1{#1}\fi
\providecommand{\url}[1]{\href{#1}{#1}}
\providecommand{\dodoi}[1]{doi:~\href{http://doi.org/#1}{\nolinkurl{#1}}}
\providecommand{\doeprint}[1]{\href{http://ascl.net/#1}{\nolinkurl{http://ascl.net/#1}}}
\providecommand{\doarXiv}[1]{\href{https://arxiv.org/abs/#1}{\nolinkurl{https://arxiv.org/abs/#1}}}

\bibitem[{{Ar{\'e}valo} {et~al.}(2025){Ar{\'e}valo}, {S{\'a}nchez-S{\'a}ez},
  {Sotomayor}, {Lira}, {Bauer}, \& {R{\'\i}os}}]{Arevalo2025arXiv251006898A}
{Ar{\'e}valo}, P., {S{\'a}nchez-S{\'a}ez}, P., {Sotomayor}, B., {et~al.} 2025,
  arXiv e-prints, arXiv:2510.06898, \dodoi{10.48550/arXiv.2510.06898}

\bibitem[{{Ar{\'e}valo} {et~al.}(2024){Ar{\'e}valo}, {L{\'o}pez-Navas},
  {Mart{\'\i}nez-Aldama}, {Lira}, {Bernal}, {S{\'a}nchez-S{\'a}ez}, {Salvato},
  {Hern{\'a}ndez-Garc{\'\i}a}, {Ricci}, {Merloni}, \&
  {Krumpe}}]{Arevalo2024A&A...683L...8A}
{Ar{\'e}valo}, P., {L{\'o}pez-Navas}, E., {Mart{\'\i}nez-Aldama}, M.~L.,
  {et~al.} 2024, \aap, 683, L8, \dodoi{10.1051/0004-6361/202348900}

\bibitem[{{Assef} {et~al.}(2011){Assef}, {Kochanek}, {Ashby}, {Brodwin},
  {Brown}, {Cool}, {Forman}, {Gonzalez}, {Hickox}, {Jannuzi}, {Jones}, {Le
  Floc'h}, {Moustakas}, {Murray}, \& {Stern}}]{Assef2011ApJ...728...56A}
{Assef}, R.~J., {Kochanek}, C.~S., {Ashby}, M.~L.~N., {et~al.} 2011, \apj, 728,
  56, \dodoi{10.1088/0004-637X/728/1/56}

\bibitem[{{Bellm} {et~al.}(2019){Bellm}, {Kulkarni}, {Graham}, {Dekany},
  {Smith}, {Riddle}, {Masci}, {Helou}, {Prince}, {Adams}, {Barbarino},
  {Barlow}, {Bauer}, {Beck}, {Belicki}, {Biswas}, {Blagorodnova}, {Bodewits},
  {Bolin}, {Brinnel}, {Brooke}, {Bue}, {Bulla}, {Burruss}, {Cenko}, {Chang},
  {Connolly}, {Coughlin}, {Cromer}, {Cunningham}, {De}, {Delacroix}, {Desai},
  {Duev}, {Eadie}, {Farnham}, {Feeney}, {Feindt}, {Flynn}, {Franckowiak},
  {Frederick}, {Fremling}, {Gal-Yam}, {Gezari}, {Giomi}, {Goldstein},
  {Golkhou}, {Goobar}, {Groom}, {Hacopians}, {Hale}, {Henning}, {Ho}, {Hover},
  {Howell}, {Hung}, {Huppenkothen}, {Imel}, {Ip}, {Ivezi{\'c}}, {Jackson},
  {Jones}, {Juric}, {Kasliwal}, {Kaspi}, {Kaye}, {Kelley}, {Kowalski},
  {Kramer}, {Kupfer}, {Landry}, {Laher}, {Lee}, {Lin}, {Lin}, {Lunnan},
  {Giomi}, {Mahabal}, {Mao}, {Miller}, {Monkewitz}, {Murphy}, {Ngeow},
  {Nordin}, {Nugent}, {Ofek}, {Patterson}, {Penprase}, {Porter}, {Rauch},
  {Rebbapragada}, {Reiley}, {Rigault}, {Rodriguez}, {van Roestel}, {Rusholme},
  {van Santen}, {Schulze}, {Shupe}, {Singer}, {Soumagnac}, {Stein}, {Surace},
  {Sollerman}, {Szkody}, {Taddia}, {Terek}, {Van Sistine}, {van Velzen},
  {Vestrand}, {Walters}, {Ward}, {Ye}, {Yu}, {Yan}, \&
  {Zolkower}}]{ZTF_2019PASP..131a8002B}
{Bellm}, E.~C., {Kulkarni}, S.~R., {Graham}, M.~J., {et~al.} 2019, \pasp, 131,
  018002, \dodoi{10.1088/1538-3873/aaecbe}

\bibitem[{{Benati Gon{\c{c}}alves} {et~al.}(2025){Benati Gon{\c{c}}alves},
  {Panda}, {Storchi Bergmann}, {Cackett}, \&
  {Eracleous}}]{Hygor_2025ApJ...988...27B}
{Benati Gon{\c{c}}alves}, H., {Panda}, S., {Storchi Bergmann}, T., {Cackett},
  E.~M., \& {Eracleous}, M. 2025, \apj, 988, 27,
  \dodoi{10.3847/1538-4357/addec0}

\bibitem[{{Bentz} {et~al.}(2009){Bentz}, {Peterson}, {Netzer}, {Pogge}, \&
  {Vestergaard}}]{Bentz_2009ApJ...697..160B}
{Bentz}, M.~C., {Peterson}, B.~M., {Netzer}, H., {Pogge}, R.~W., \&
  {Vestergaard}, M. 2009, \apj, 697, 160, \dodoi{10.1088/0004-637X/697/1/160}

\bibitem[{{Bernal} {et~al.}(2025){Bernal}, {S{\'a}nchez-S{\'a}ez},
  {Ar{\'e}valo}, {Bauer}, {Lira}, \& {Sotomayor}}]{Bernal2025A&A...694A.127B}
{Bernal}, S., {S{\'a}nchez-S{\'a}ez}, P., {Ar{\'e}valo}, P., {et~al.} 2025,
  \aap, 694, A127, \dodoi{10.1051/0004-6361/202451870}

\bibitem[{{Besuner} {et~al.}(2025){Besuner}, {Dey}, {Drlica-Wagner}, {Ebina},
  {Fernandez Moroni}, {Ferraro}, {Forero-Romero}, {Honscheid}, {Jelinsky},
  {Lang}, {Levi}, {Martini}, {Myers}, {Palanque-Delabrouille}, {Panda},
  {Poppett}, {Sailer}, {Schlegel}, {Shafieloo}, {Silber}, {White}, {Abbott},
  {Allen}, {Avila}, {Avil{\'e}s}, {Bailey}, {Bault}, {Bouri}, {Boutsia},
  {Burtin}, {Chierchie}, {Coulton}, {Dawson}, {Dey}, {Dor{\'e}}, {Dunlop},
  {Eisenstein}, {Emanuele}, {Escoffier}, {Estrada}, {Fagrelius}, {Fanning},
  {Fanning}, {Font-Ribera}, {Frieman}, {Galal}, {Gluscevic}, {Gontcho},
  {Green}, {Gutierrez}, {Guy}, {Hashemi}, {Heathcote}, {Holland}, {Hou},
  {Huterer}, {Irigoyen Gimenez}, {Ivanov}, {Joyce}, {Jullo}, {Juneau},
  {Juramy}, {Karcher}, {Kent}, {Kirkby}, {Kneib}, {Krause}, {Krolewski},
  {Lahav}, {Lapi}, {Leauthaud}, {Lewandowski}, {Li}, {Lin}, {Loverde},
  {MacBride}, {Magneville}, {Marshall}, {McDonald}, {Miller}, {Moustakas},
  {M{\"u}nchmeyer}, {Najita}, {Newman}, {Percival}, {Philcox}, {Pires},
  {Raichoor}, {Roach}, {Rockosi}, {Rombach}, {Ross}, {Sanchez}, {Schmidt},
  {Schubnell}, {Sebok}, {Seljak}, {Silverstein}, {Slepian}, {Stone}, {Stupak},
  {Tarl{\'e}}, {Li}, {Tyas}, {Vargas-Maga{\~n}a}, {Walker}, {Wenner},
  {Y{\`e}che}, {Zhang}, \& {Zhou}}]{specs5_2025arXiv250307923B}
{Besuner}, R., {Dey}, A., {Drlica-Wagner}, A., {et~al.} 2025, arXiv e-prints,
  arXiv:2503.07923, \dodoi{10.48550/arXiv.2503.07923}

\bibitem[{{Blandford} \& {McKee}(1982)}]{Blandford_1982ApJ...255..419B}
{Blandford}, R.~D., \& {McKee}, C.~F. 1982, \apj, 255, 419,
  \dodoi{10.1086/159843}

\bibitem[{{Bon} {et~al.}(2016){Bon}, {Zucker}, {Netzer}, {Marziani}, {Bon},
  {Jovanovi{\'c}}, {Shapovalova}, {Komossa}, {Gaskell}, {Popovi{\'c}},
  {Britzen}, {Chavushyan}, {Burenkov}, {Sergeev}, {La Mura}, {Vald{\'e}s}, \&
  {Stalevski}}]{Bon2016ApJS..225...29B}
{Bon}, E., {Zucker}, S., {Netzer}, H., {et~al.} 2016, \apjs, 225, 29,
  \dodoi{10.3847/0067-0049/225/2/29}

\bibitem[{{Brandt} {et~al.}(2018){Brandt}, {Ni}, {Yang}, {Anderson}, {Assef},
  {Barth}, {Bauer}, {Bongiorno}, {Chen}, {De Cicco}, {Gezari}, {Grier}, {Hall},
  {Hoenig}, {Lacy}, {Li}, {Luo}, {Paolillo}, {Peterson}, {Popovi{\'c}},
  {Richards}, {Shemmer}, {Shen}, {Sun}, {Timlin}, {Trump}, {Vito}, \&
  {Yu}}]{Brandt2018arXiv181106542B}
{Brandt}, W.~N., {Ni}, Q., {Yang}, G., {et~al.} 2018, arXiv e-prints,
  arXiv:1811.06542, \dodoi{10.48550/arXiv.1811.06542}

\bibitem[{{Cackett} {et~al.}(2021){Cackett}, {Bentz}, \&
  {Kara}}]{Cackett_2021iSci...24j2557C}
{Cackett}, E.~M., {Bentz}, M.~C., \& {Kara}, E. 2021, iScience, 24, 102557,
  \dodoi{10.1016/j.isci.2021.102557}

\bibitem[{{Cackett} {et~al.}(2018){Cackett}, {Chiang}, {McHardy}, {Edelson},
  {Goad}, {Horne}, \& {Korista}}]{Cackett2018ApJ...857...53C}
{Cackett}, E.~M., {Chiang}, C.-Y., {McHardy}, I., {et~al.} 2018, \apj, 857, 53,
  \dodoi{10.3847/1538-4357/aab4f7}

\bibitem[{{Charisi} {et~al.}(2022){Charisi}, {Taylor}, {Runnoe}, {Bogdanovic},
  \& {Trump}}]{Charisi2022MNRAS.510.5929C}
{Charisi}, M., {Taylor}, S.~R., {Runnoe}, J., {Bogdanovic}, T., \& {Trump},
  J.~R. 2022, \mnras, 510, 5929, \dodoi{10.1093/mnras/stab3713}

\bibitem[{{Chelouche} {et~al.}(2019){Chelouche}, {Pozo Nu{\~n}ez}, \&
  {Kaspi}}]{Chelouche2019NatAs...3..251C}
{Chelouche}, D., {Pozo Nu{\~n}ez}, F., \& {Kaspi}, S. 2019, Nature Astronomy,
  3, 251, \dodoi{10.1038/s41550-018-0659-x}

\bibitem[{{Czerny} {et~al.}(2023){Czerny}, {Panda}, {Prince}, {Kumar Jaiswal},
  {Zaja{\v{c}}ek}, {Martinez Aldama}, {Koz{\l}owski}, {Kovacevic}, {Ilic},
  {Popovi{\'c}}, {Pozo Nu{\~n}ez}, {H{\"o}nig}, \&
  {Brandt}}]{Czerny_2023A&A...675A.163C}
{Czerny}, B., {Panda}, S., {Prince}, R., {et~al.} 2023, \aap, 675, A163,
  \dodoi{10.1051/0004-6361/202345844}

\bibitem[{{Dalla Bont{\`a}} {et~al.}(2020){Dalla Bont{\`a}}, {Peterson},
  {Bentz}, {Brandt}, {Ciroi}, {De Rosa}, {Fonseca Alvarez}, {Grier}, {Hall},
  {Hern{\'a}ndez Santisteban}, {Ho}, {Homayouni}, {Horne}, {Kochanek}, {Li},
  {Morelli}, {Pizzella}, {Pogge}, {Schneider}, {Shen}, {Trump}, \&
  {Vestergaard}}]{DallaBonta2020ApJ...903..112D}
{Dalla Bont{\`a}}, E., {Peterson}, B.~M., {Bentz}, M.~C., {et~al.} 2020, \apj,
  903, 112, \dodoi{10.3847/1538-4357/abbc1c}

\bibitem[{{D'Orazio} \& {Charisi}(2023)}]{DOrazio2023arXiv231016896D}
{D'Orazio}, D.~J., \& {Charisi}, M. 2023, arXiv e-prints, arXiv:2310.16896,
  \dodoi{10.48550/arXiv.2310.16896}

\bibitem[{{Edelson} {et~al.}(2017){Edelson}, {Gelbord}, {Cackett}, {Connolly},
  {Done}, {Fausnaugh}, {Gardner}, {Gehrels}, {Goad}, {Horne}, {McHardy},
  {Peterson}, {Vaughan}, {Vestergaard}, {Breeveld}, {Barth}, {Bentz},
  {Bottorff}, {Brandt}, {Crawford}, {Dalla Bont{\`a}}, {Emmanoulopoulos},
  {Evans}, {Figuera Jaimes}, {Filippenko}, {Ferland}, {Grupe}, {Joner},
  {Kennea}, {Korista}, {Krimm}, {Kriss}, {Leonard}, {Mathur}, {Netzer},
  {Nousek}, {Page}, {Romero-Colmenero}, {Siegel}, {Starkey}, {Treu}, {Vogler},
  {Winkler}, \& {Zheng}}]{Edelson2017ApJ...840...41E}
{Edelson}, R., {Gelbord}, J., {Cackett}, E., {et~al.} 2017, \apj, 840, 41,
  \dodoi{10.3847/1538-4357/aa6890}

\bibitem[{{El-Badry} {et~al.}(2025){El-Badry}, {Hogg}, \&
  {Rix}}]{ElBadry2025arXiv250910601E}
{El-Badry}, K., {Hogg}, D.~W., \& {Rix}, H.-W. 2025, arXiv e-prints,
  arXiv:2509.10601, \dodoi{10.48550/arXiv.2509.10601}

\bibitem[{{Fan} {et~al.}(2023){Fan}, {Ba{\~n}ados}, \&
  {Simcoe}}]{Fan2023ARA&A..61..373F}
{Fan}, X., {Ba{\~n}ados}, E., \& {Simcoe}, R.~A. 2023, \araa, 61, 373,
  \dodoi{10.1146/annurev-astro-052920-102455}

\bibitem[{{Fan} {et~al.}(2006){Fan}, {Carilli}, \&
  {Keating}}]{Fan2006ARA&A..44..415F}
{Fan}, X., {Carilli}, C.~L., \& {Keating}, B. 2006, \araa, 44, 415,
  \dodoi{10.1146/annurev.astro.44.051905.092514}

\bibitem[{{Fatovi{\'c}} {et~al.}(2025){Fatovi{\'c}}, {Ili{\'c}},
  {Kova{\v{c}}evi{\'c}}, {Palaversa}, {Simi{\'c}}, {Popovi{\'c}}, {Thanjavur},
  {Razim}, {Ivezi{\'c}}, {Yue}, \& {Fan}}]{Fatovic2025A&A...695A.208F}
{Fatovi{\'c}}, M., {Ili{\'c}}, D., {Kova{\v{c}}evi{\'c}}, A.~B., {et~al.} 2025,
  \aap, 695, A208, \dodoi{10.1051/0004-6361/202453600}

\bibitem[{{Fausnaugh} {et~al.}(2016){Fausnaugh}, {Denney}, {Barth}, {Bentz},
  {Bottorff}, {Carini}, {Croxall}, {De Rosa}, {Goad}, {Horne}, {Joner},
  {Kaspi}, {Kim}, {Klimanov}, {Kochanek}, {Leonard}, {Netzer}, {Peterson},
  {Schn{\"u}lle}, {Sergeev}, {Vestergaard}, {Zheng}, {Zu}, {Anderson},
  {Ar{\'e}valo}, {Bazhaw}, {Borman}, {Boroson}, {Brandt}, {Breeveld}, {Brewer},
  {Cackett}, {Crenshaw}, {Dalla Bont{\`a}}, {De Lorenzo-C{\'a}ceres},
  {Dietrich}, {Edelson}, {Efimova}, {Ely}, {Evans}, {Filippenko}, {Flatland},
  {Gehrels}, {Geier}, {Gelbord}, {Gonzalez}, {Gorjian}, {Grier}, {Grupe},
  {Hall}, {Hicks}, {Horenstein}, {Hutchison}, {Im}, {Jensen}, {Jones},
  {Kaastra}, {Kelly}, {Kennea}, {Kim}, {Korista}, {Kriss}, {Lee}, {Lira},
  {MacInnis}, {Manne-Nicholas}, {Mathur}, {McHardy}, {Montouri}, {Musso},
  {Nazarov}, {Norris}, {Nousek}, {Okhmat}, {Pancoast}, {Papadakis}, {Parks},
  {Pei}, {Pogge}, {Pott}, {Rafter}, {Rix}, {Saylor}, {Schimoia}, {Siegel},
  {Spencer}, {Starkey}, {Sung}, {Teems}, {Treu}, {Turner}, {Uttley},
  {Villforth}, {Weiss}, {Woo}, {Yan}, \&
  {Young}}]{Fausnaugh_2016ApJ...821...56F}
{Fausnaugh}, M.~M., {Denney}, K.~D., {Barth}, A.~J., {et~al.} 2016, \apj, 821,
  56, \dodoi{10.3847/0004-637X/821/1/56}

\bibitem[{{Fiore} {et~al.}(2017){Fiore}, {Feruglio}, {Shankar}, {Bischetti},
  {Bongiorno}, {Brusa}, {Carniani}, {Cicone}, {Duras}, {Lamastra}, {Mainieri},
  {Marconi}, {Menci}, {Maiolino}, {Piconcelli}, {Vietri}, \&
  {Zappacosta}}]{Fiore2017A&A...601A.143F}
{Fiore}, F., {Feruglio}, C., {Shankar}, F., {et~al.} 2017, \aap, 601, A143,
  \dodoi{10.1051/0004-6361/201629478}

\bibitem[{{Frederick} {et~al.}(2021){Frederick}, {Gezari}, {Graham},
  {Sollerman}, {van Velzen}, {Perley}, {Stern}, {Ward}, {Hammerstein}, {Hung},
  {Yan}, {Andreoni}, {Bellm}, {Duev}, {Kowalski}, {Mahabal}, {Masci},
  {Medford}, {Rusholme}, {Smith}, \& {Walters}}]{Frederick2021ApJ...920...56F}
{Frederick}, S., {Gezari}, S., {Graham}, M.~J., {et~al.} 2021, \apj, 920, 56,
  \dodoi{10.3847/1538-4357/ac110f}

\bibitem[{{Frohmaier} {et~al.}(2025){Frohmaier}, {Vincenzi}, {Sullivan},
  {H{\"o}nig}, {Smith}, {Addison}, {Collett}, {Dimitriadis}, {Ellis}, {Gandhi},
  {Graur}, {Hook}, {Kelsey}, {Kim}, {Lidman}, {Maguire}, {Makrygianni},
  {Martin}, {M{\"o}ller}, {Nichol}, {Nicholl}, {Schady}, {Simmons}, {Smartt},
  {Tempel}, {Wiseman}, \& {the LSST Dark Energy Science
  Collaboration}}]{4MOST_2025ApJ...992..158F}
{Frohmaier}, C., {Vincenzi}, M., {Sullivan}, M., {et~al.} 2025, \apj, 992, 158,
  \dodoi{10.3847/1538-4357/adff4e}

\bibitem[{{Gardner} {et~al.}(2006){Gardner}, {Mather}, {Clampin}, {Doyon},
  {Greenhouse}, {Hammel}, {Hutchings}, {Jakobsen}, {Lilly}, {Long}, {Lunine},
  {McCaughrean}, {Mountain}, {Nella}, {Rieke}, {Rieke}, {Rix}, {Smith},
  {Sonneborn}, {Stiavelli}, {Stockman}, {Windhorst}, \&
  {Wright}}]{JWST_2006SSRv..123..485G}
{Gardner}, J.~P., {Mather}, J.~C., {Clampin}, M., {et~al.} 2006, \ssr, 123,
  485, \dodoi{10.1007/s11214-006-8315-7}

\bibitem[{{Gezari}(2021)}]{Gezari2021ARA&A..59...21G}
{Gezari}, S. 2021, \araa, 59, 21, \dodoi{10.1146/annurev-astro-111720-030029}

\bibitem[{{Graham} {et~al.}(2019){Graham}, {Kulkarni}, {Bellm}, {Adams},
  {Barbarino}, {Blagorodnova}, {Bodewits}, {Bolin}, {Brady}, {Cenko}, {Chang},
  {Coughlin}, {De}, {Eadie}, {Farnham}, {Feindt}, {Franckowiak}, {Fremling},
  {Gezari}, {Ghosh}, {Goldstein}, {Golkhou}, {Goobar}, {Ho}, {Huppenkothen},
  {Ivezi{\'c}}, {Jones}, {Juric}, {Kaplan}, {Kasliwal}, {Kelley}, {Kupfer},
  {Lee}, {Lin}, {Lunnan}, {Mahabal}, {Miller}, {Ngeow}, {Nugent}, {Ofek},
  {Prince}, {Rauch}, {van Roestel}, {Schulze}, {Singer}, {Sollerman}, {Taddia},
  {Yan}, {Ye}, {Yu}, {Barlow}, {Bauer}, {Beck}, {Belicki}, {Biswas}, {Brinnel},
  {Brooke}, {Bue}, {Bulla}, {Burruss}, {Connolly}, {Cromer}, {Cunningham},
  {Dekany}, {Delacroix}, {Desai}, {Duev}, {Feeney}, {Flynn}, {Frederick},
  {Gal-Yam}, {Giomi}, {Groom}, {Hacopians}, {Hale}, {Helou}, {Henning},
  {Hover}, {Hillenbrand}, {Howell}, {Hung}, {Imel}, {Ip}, {Jackson}, {Kaspi},
  {Kaye}, {Kowalski}, {Kramer}, {Kuhn}, {Landry}, {Laher}, {Mao}, {Masci},
  {Monkewitz}, {Murphy}, {Nordin}, {Patterson}, {Penprase}, {Porter},
  {Rebbapragada}, {Reiley}, {Riddle}, {Rigault}, {Rodriguez}, {Rusholme}, {van
  Santen}, {Shupe}, {Smith}, {Soumagnac}, {Stein}, {Surace}, {Szkody}, {Terek},
  {Van Sistine}, {van Velzen}, {Vestrand}, {Walters}, {Ward}, {Zhang}, \&
  {Zolkower}}]{ZTF_2019PASP..131g8001G}
{Graham}, M.~J., {Kulkarni}, S.~R., {Bellm}, E.~C., {et~al.} 2019, \pasp, 131,
  078001, \dodoi{10.1088/1538-3873/ab006c}

\bibitem[{{Graham} {et~al.}(2020){Graham}, {Ross}, {Stern}, {Drake},
  {McKernan}, {Ford}, {Djorgovski}, {Mahabal}, {Glikman}, {Larson}, \&
  {Christensen}}]{Graham_2020MNRAS.491.4925G}
{Graham}, M.~J., {Ross}, N.~P., {Stern}, D., {et~al.} 2020, \mnras, 491, 4925,
  \dodoi{10.1093/mnras/stz3244}

\bibitem[{{Graham} {et~al.}(2023){Graham}, {McKernan}, {Ford}, {Stern},
  {Djorgovski}, {Coughlin}, {Burdge}, {Bellm}, {Helou}, {Mahabal}, {Masci},
  {Purdum}, {Rosnet}, \& {Rusholme}}]{Graham2023ApJ...942...99G}
{Graham}, M.~J., {McKernan}, B., {Ford}, K.~E.~S., {et~al.} 2023, \apj, 942,
  99, \dodoi{10.3847/1538-4357/aca480}

\bibitem[{{Greene} {et~al.}(2020){Greene}, {Strader}, \&
  {Ho}}]{Greene2020ARA&A..58..257G}
{Greene}, J.~E., {Strader}, J., \& {Ho}, L.~C. 2020, \araa, 58, 257,
  \dodoi{10.1146/annurev-astro-032620-021835}

\bibitem[{{Guo} {et~al.}(2022{\natexlab{a}}){Guo}, {Barth}, \&
  {Wang}}]{Guo2022ApJ...940...20G}
{Guo}, H., {Barth}, A.~J., \& {Wang}, S. 2022{\natexlab{a}}, \apj, 940, 20,
  \dodoi{10.3847/1538-4357/ac96ec}

\bibitem[{{Guo} {et~al.}(2022{\natexlab{b}}){Guo}, {Li}, {Zhang}, {Ho}, \&
  {Wang}}]{Guo2022ApJ...929...19G}
{Guo}, W.-J., {Li}, Y.-R., {Zhang}, Z.-X., {Ho}, L.~C., \& {Wang}, J.-M.
  2022{\natexlab{b}}, \apj, 929, 19, \dodoi{10.3847/1538-4357/ac4e84}

\bibitem[{{Guo} {et~al.}(2025{\natexlab{a}}){Guo}, {Fawcett}, {Siudek}, {Li},
  {Cheng}, {Panda}, {Pan}, {Sun}, {Greenwell}, {Alexander}, {Moustakas},
  {Zhai}, {Jin}, {Cheng}, {Hu}, {Chen}, {Zhang}, \&
  {Wang}}]{Guo_2025ApJ...995..139G}
{Guo}, W.-J., {Fawcett}, V.~A., {Siudek}, M., {et~al.} 2025{\natexlab{a}},
  \apj, 995, 139, \dodoi{10.3847/1538-4357/ae1d7b}

\bibitem[{{Guo} {et~al.}(2025{\natexlab{b}}){Guo}, {Zou}, {Greenwell},
  {Alexander}, {Fawcett}, {Pan}, {Siudek}, {Aguilar}, {Ahlen}, {Brooks},
  {Claybaugh}, {Dawson}, {de la Macorra}, {Doel}, {Font-Ribera},
  {Gazta{\~n}aga}, {Gontcho A Gontcho}, {Gutierrez}, {Kehoe}, {Kisner},
  {Landriau}, {Le Guillou}, {Manera}, {Meisner}, {Miquel}, {Moustakas},
  {Prada}, {Rossi}, {Sanchez}, {Schubnell}, {Sprayberry}, {Sui}, {Tarl{\'e}},
  {Weaver}, {Xiao}, \& {Zou}}]{Guo_DESI_2025ApJS..278...28G}
{Guo}, W.-J., {Zou}, H., {Greenwell}, C.~L., {et~al.} 2025{\natexlab{b}},
  \apjs, 278, 28, \dodoi{10.3847/1538-4365/adc124}

\bibitem[{{Gupta} {et~al.}(2025){Gupta}, {Muthukrishna}, {Rehemtulla}, \&
  {Shah}}]{Gupta2025MNRAS.542L.132G}
{Gupta}, R., {Muthukrishna}, D., {Rehemtulla}, N., \& {Shah}, V. 2025, \mnras,
  542, L132, \dodoi{10.1093/mnrasl/slaf074}

\bibitem[{{Hagen} {et~al.}(2024){Hagen}, {Done}, \&
  {Edelson}}]{Hagen_2024MNRAS.530.4850H}
{Hagen}, S., {Done}, C., \& {Edelson}, R. 2024, \mnras, 530, 4850,
  \dodoi{10.1093/mnras/stae1177}

\bibitem[{{Hern{\'a}ndez-Garc{\'\i}a}
  {et~al.}(2025){Hern{\'a}ndez-Garc{\'\i}a}, {Chakraborty},
  {S{\'a}nchez-S{\'a}ez}, {Ricci}, {Cuadra}, {McKernan}, {Ford}, {Ar{\'e}valo},
  {Rau}, {Arcodia}, {Kara}, {Liu}, {Merloni}, {Bruni}, {Goodwin},
  {Arzoumanian}, {Assef}, {Baldini}, {Bayo}, {Bauer}, {Bernal}, {Brightman},
  {Calistro Rivera}, {Gendreau}, {Homan}, {Krumpe}, {Lira},
  {Mart{\'\i}nez-Aldama}, {Salvato}, \&
  {Sotomayor}}]{Hernandez_Garcia2025NatAs...9..895H}
{Hern{\'a}ndez-Garc{\'\i}a}, L., {Chakraborty}, J., {S{\'a}nchez-S{\'a}ez}, P.,
  {et~al.} 2025, Nature Astronomy, 9, 895, \dodoi{10.1038/s41550-025-02523-9}

\bibitem[{{Hinkle}(2024)}]{Hinkle2024MNRAS.531.2603H}
{Hinkle}, J.~T. 2024, \mnras, 531, 2603, \dodoi{10.1093/mnras/stae1229}

\bibitem[{{Holoien} {et~al.}(2022){Holoien}, {Neustadt}, {Vallely}, {Auchettl},
  {Hinkle}, {Romero-Ca{\~n}izales}, {Shappee}, {Kochanek}, {Stanek}, {Chen},
  {Dong}, {Prieto}, {Thompson}, {Brink}, {Filippenko}, {Zheng}, {Bersier},
  {Bose}, {Burgasser}, {Channa}, {de Jaeger}, {Hestenes}, {Im}, {Jeffers},
  {Jun}, {Lansbury}, {Post}, {Ross}, {Stern}, {Tang}, {Tucker}, {Valenti},
  {Yunus}, \& {Zhang}}]{Holoien2022ApJ...933..196H}
{Holoien}, T. W.-S., {Neustadt}, J. M.~M., {Vallely}, P.~J., {et~al.} 2022,
  \apj, 933, 196, \dodoi{10.3847/1538-4357/ac74b9}

\bibitem[{{Homayouni} {et~al.}(2019){Homayouni}, {Trump}, {Grier}, {Shen},
  {Starkey}, {Brandt}, {Fonseca Alvarez}, {Hall}, {Horne}, {Kinemuchi}, {I-Hsiu
  Li}, {McGreer}, {Sun}, {Ho}, \& {Schneider}}]{Homayouni2019ApJ...880..126H}
{Homayouni}, Y., {Trump}, J.~R., {Grier}, C.~J., {et~al.} 2019, \apj, 880, 126,
  \dodoi{10.3847/1538-4357/ab2638}

\bibitem[{{Hopkins} {et~al.}(2007){Hopkins}, {Richards}, \&
  {Hernquist}}]{Hopkins2007ApJ...654..731H}
{Hopkins}, P.~F., {Richards}, G.~T., \& {Hernquist}, L. 2007, \apj, 654, 731,
  \dodoi{10.1086/509629}

\bibitem[{{Horne} {et~al.}(2004){Horne}, {Peterson}, {Collier}, \&
  {Netzer}}]{Horne_2004PASP..116..465H}
{Horne}, K., {Peterson}, B.~M., {Collier}, S.~J., \& {Netzer}, H. 2004, \pasp,
  116, 465, \dodoi{10.1086/420755}

\bibitem[{{Ivezi{\'c}} {et~al.}(2019){Ivezi{\'c}}, {Kahn}, {Tyson}, {Abel},
  {Acosta}, {Allsman}, {Alonso}, {AlSayyad}, {Anderson}, {Andrew}, {Angel},
  {Angeli}, {Ansari}, {Antilogus}, {Araujo}, {Armstrong}, {Arndt}, {Astier},
  {Aubourg}, {Auza}, {Axelrod}, {Bard}, {Barr}, {Barrau}, {Bartlett}, {Bauer},
  {Bauman}, {Baumont}, {Bechtol}, {Bechtol}, {Becker}, {Becla}, {Beldica},
  {Bellavia}, {Bianco}, {Biswas}, {Blanc}, {Blazek}, {Blandford}, {Bloom},
  {Bogart}, {Bond}, {Booth}, {Borgland}, {Borne}, {Bosch}, {Boutigny},
  {Brackett}, {Bradshaw}, {Brandt}, {Brown}, {Bullock}, {Burchat}, {Burke},
  {Cagnoli}, {Calabrese}, {Callahan}, {Callen}, {Carlin}, {Carlson},
  {Chandrasekharan}, {Charles-Emerson}, {Chesley}, {Cheu}, {Chiang}, {Chiang},
  {Chirino}, {Chow}, {Ciardi}, {Claver}, {Cohen-Tanugi}, {Cockrum}, {Coles},
  {Connolly}, {Cook}, {Cooray}, {Covey}, {Cribbs}, {Cui}, {Cutri}, {Daly},
  {Daniel}, {Daruich}, {Daubard}, {Daues}, {Dawson}, {Delgado}, {Dellapenna},
  {de Peyster}, {de Val-Borro}, {Digel}, {Doherty}, {Dubois},
  {Dubois-Felsmann}, {Durech}, {Economou}, {Eifler}, {Eracleous}, {Emmons},
  {Fausti Neto}, {Ferguson}, {Figueroa}, {Fisher-Levine}, {Focke}, {Foss},
  {Frank}, {Freemon}, {Gangler}, {Gawiser}, {Geary}, {Gee}, {Geha}, {Gessner},
  {Gibson}, {Gilmore}, {Glanzman}, {Glick}, {Goldina}, {Goldstein}, {Goodenow},
  {Graham}, {Gressler}, {Gris}, {Guy}, {Guyonnet}, {Haller}, {Harris},
  {Hascall}, {Haupt}, {Hernandez}, {Herrmann}, {Hileman}, {Hoblitt}, {Hodgson},
  {Hogan}, {Howard}, {Huang}, {Huffer}, {Ingraham}, {Innes}, {Jacoby}, {Jain},
  {Jammes}, {Jee}, {Jenness}, {Jernigan}, {Jevremovi{\'c}}, {Johns}, {Johnson},
  {Johnson}, {Jones}, {Juramy-Gilles}, {Juri{\'c}}, {Kalirai}, {Kallivayalil},
  {Kalmbach}, {Kantor}, {Karst}, {Kasliwal}, {Kelly}, {Kessler}, {Kinnison},
  {Kirkby}, {Knox}, {Kotov}, {Krabbendam}, {Krughoff}, {Kub{\'a}nek},
  {Kuczewski}, {Kulkarni}, {Ku}, {Kurita}, {Lage}, {Lambert}, {Lange},
  {Langton}, {Le Guillou}, {Levine}, {Liang}, {Lim}, {Lintott}, {Long},
  {Lopez}, {Lotz}, {Lupton}, {Lust}, {MacArthur}, {Mahabal}, {Mandelbaum},
  {Markiewicz}, {Marsh}, {Marshall}, {Marshall}, {May}, {McKercher}, {McQueen},
  {Meyers}, {Migliore}, {Miller}, \& {Mills}}]{LSST_2019ApJ...873..111I}
{Ivezi{\'c}}, {\v{Z}}., {Kahn}, S.~M., {Tyson}, J.~A., {et~al.} 2019, \apj,
  873, 111, \dodoi{10.3847/1538-4357/ab042c}

\bibitem[{{Jaiswal} {et~al.}(2025){Jaiswal}, {Mandal}, {Prince}, {Pandey},
  {Naddaf}, {Czerny}, {Panda}, \& {Pozo
  Nu{\~n}ez}}]{Jaiswal2025A&A...702A..92J}
{Jaiswal}, V.~K., {Mandal}, A.~K., {Prince}, R., {et~al.} 2025, \aap, 702, A92,
  \dodoi{10.1051/0004-6361/202452497}

\bibitem[{{Jana} {et~al.}(2025){Jana}, {Ricci}, {Temple}, {Chang},
  {Shablovinskaya}, {Trakhtenbrot}, {Diaz}, {Ilic}, {Nandi}, \&
  {Koss}}]{Jana_2025A&A...693A..35J}
{Jana}, A., {Ricci}, C., {Temple}, M.~J., {et~al.} 2025, \aap, 693, A35,
  \dodoi{10.1051/0004-6361/202451058}

\bibitem[{{Jha} {et~al.}(2022){Jha}, {Joshi}, {Chand}, {Wu}, {Ho}, {Rastogi},
  \& {Ma}}]{Jha2022MNRAS.511.3005J}
{Jha}, V.~K., {Joshi}, R., {Chand}, H., {et~al.} 2022, \mnras, 511, 3005,
  \dodoi{10.1093/mnras/stac109}

\bibitem[{{Jones} {et~al.}(2014){Jones}, {Yoachim}, {Chandrasekharan},
  {Connolly}, {Cook}, {Ivezic}, {Krughoff}, {Petry}, \&
  {Ridgway}}]{Jones2014SPIE.9149E..0BJ}
{Jones}, R.~L., {Yoachim}, P., {Chandrasekharan}, S., {et~al.} 2014, in Society
  of Photo-Optical Instrumentation Engineers (SPIE) Conference Series, Vol.
  9149, Observatory Operations: Strategies, Processes, and Systems V, ed. A.~B.
  {Peck}, C.~R. {Benn}, \& R.~L. {Seaman}, 91490B, \dodoi{10.1117/12.2056835}

\bibitem[{{Ju} {et~al.}(2013){Ju}, {Greene}, {Rafikov}, {Bickerton}, \&
  {Badenes}}]{Ju2013ApJ...777...44J}
{Ju}, W., {Greene}, J.~E., {Rafikov}, R.~R., {Bickerton}, S.~J., \& {Badenes},
  C. 2013, \apj, 777, 44, \dodoi{10.1088/0004-637X/777/1/44}

\bibitem[{{Kaspi} {et~al.}(2021){Kaspi}, {Brandt}, {Maoz}, {Netzer},
  {Schneider}, {Shemmer}, \& {Grier}}]{Kaspi_2021ApJ...915..129K}
{Kaspi}, S., {Brandt}, W.~N., {Maoz}, D., {et~al.} 2021, \apj, 915, 129,
  \dodoi{10.3847/1538-4357/ac00aa}

\bibitem[{{Komossa} {et~al.}(2024){Komossa}, {Grupe}, {Marziani}, {Popovic},
  {Marceta-Mandic}, {Bon}, {Ilic}, {Kovacevic}, {Kraus}, {Haiman}, {Petrecca},
  {De Cicco}, {Dimitrijevic}, {Sreckovic}, {Kovacevic Dojcinovic},
  {Pannikkote}, {Bon}, {Gupta}, \& {Iacob}}]{Komossa2024arXiv240800089K}
{Komossa}, S., {Grupe}, D., {Marziani}, P., {et~al.} 2024, arXiv e-prints,
  arXiv:2408.00089, \dodoi{10.48550/arXiv.2408.00089}

\bibitem[{{Korista} \& {Goad}(2001)}]{Korista2001ApJ...553..695K}
{Korista}, K.~T., \& {Goad}, M.~R. 2001, \apj, 553, 695, \dodoi{10.1086/320964}

\bibitem[{{Korista} \& {Goad}(2019)}]{Korista2019MNRAS.489.5284K}
---. 2019, \mnras, 489, 5284, \dodoi{10.1093/mnras/stz2330}

\bibitem[{{Kova{\v{c}}evi{\'c}} {et~al.}(2022){Kova{\v{c}}evi{\'c}},
  {Radovi{\'c}}, {Ili{\'c}}, {Popovi{\'c}}, {Assef}, {S{\'a}nchez-S{\'a}ez},
  {Nikutta}, {Raiteri}, {Yoon}, {Homayouni}, {Li}, {Caplar}, {Czerny}, {Panda},
  {Ricci}, {Jankov}, {Landt}, {Wolf},
  {Kova{\v{c}}evi{\'c}-Doj{\v{c}}inovi{\'c}}, {Laki{\'c}evi{\'c}}, {Savi{\'c}},
  {Vince}, {Simi{\'c}}, {{\v{C}}vorovi{\'c}-Hajdinjak}, \&
  {Mar{\v{c}}eta-Mandi{\'c}}}]{Kovacevic_2022ApJS..262...49K}
{Kova{\v{c}}evi{\'c}}, A.~B., {Radovi{\'c}}, V., {Ili{\'c}}, D., {et~al.} 2022,
  \apjs, 262, 49, \dodoi{10.3847/1538-4365/ac88ce}

\bibitem[{{LaMassa} {et~al.}(2015){LaMassa}, {Cales}, {Moran}, {Myers},
  {Richards}, {Eracleous}, {Heckman}, {Gallo}, \&
  {Urry}}]{LaMassa_2015ApJ...800..144L}
{LaMassa}, S.~M., {Cales}, S., {Moran}, E.~C., {et~al.} 2015, \apj, 800, 144,
  \dodoi{10.1088/0004-637X/800/2/144}

\bibitem[{{Lawther} {et~al.}(2018){Lawther}, {Goad}, {Korista}, {Ulrich}, \&
  {Vestergaard}}]{Lawther2018MNRAS.481..533L}
{Lawther}, D., {Goad}, M.~R., {Korista}, K.~T., {Ulrich}, O., \& {Vestergaard},
  M. 2018, \mnras, 481, 533, \dodoi{10.1093/mnras/sty2242}

\bibitem[{{Li} {et~al.}(2025{\natexlab{a}}){Li}, {Assef}, {Brandt}, {Temple},
  {Bauer}, {Marculewicz}, {Panda}, {Peca}, {Ricci}, {Richards}, {Satheesh
  Sheeba}, {Tsai}, {Wu}, \& {Yoon}}]{Li_2025arXiv251208654L}
{Li}, G., {Assef}, R.~J., {Brandt}, W.~N., {et~al.} 2025{\natexlab{a}}, arXiv
  e-prints, arXiv:2512.08654, \dodoi{10.48550/arXiv.2512.08654}

\bibitem[{{Li} {et~al.}(2025{\natexlab{b}}){Li}, {Shangguan}, {Wang}, {Davies},
  {Santos}, {Eisenhauer}, {Songsheng}, {Winkler}, {Aceituno}, {Bai}, {Bai},
  {Brotherton}, {Cao}, {Chen}, {Du}, {Fang}, {Feng}, {Feuchtgruber},
  {F{\"o}rster Schreiber}, {Fu}, {Genzel}, {Gillessen}, {Ho}, {Hu}, {Liu},
  {Lutz}, {Ott}, {Petrov}, {Rabien}, {Shimizu}, {Sturm}, {Tacconi}, {Wang},
  {Yao}, {Zhai}, {Zhang}, {Zhao}, {Zhao}, \& {SARM
  Collaboration}}]{Li2025ApJ...988...42L}
{Li}, Y.-R., {Shangguan}, J., {Wang}, J.-M., {et~al.} 2025{\natexlab{b}}, \apj,
  988, 42, \dodoi{10.3847/1538-4357/addf40}

\bibitem[{{LSST Science Collaboration} {et~al.}(2009){LSST Science
  Collaboration}, {Abell}, {Allison}, {Anderson}, {Andrew}, {Angel}, {Armus},
  {Arnett}, {Asztalos}, {Axelrod}, {Bailey}, {Ballantyne}, {Bankert},
  {Barkhouse}, {Barr}, {Barrientos}, {Barth}, {Bartlett}, {Becker}, {Becla},
  {Beers}, {Bernstein}, {Biswas}, {Blanton}, {Bloom}, {Bochanski}, {Boeshaar},
  {Borne}, {Bradac}, {Brandt}, {Bridge}, {Brown}, {Brunner}, {Bullock},
  {Burgasser}, {Burge}, {Burke}, {Cargile}, {Chandrasekharan}, {Chartas},
  {Chesley}, {Chu}, {Cinabro}, {Claire}, {Claver}, {Clowe}, {Connolly}, {Cook},
  {Cooke}, {Cooray}, {Covey}, {Culliton}, {de Jong}, {de Vries}, {Debattista},
  {Delgado}, {Dell'Antonio}, {Dhital}, {Di Stefano}, {Dickinson}, {Dilday},
  {Djorgovski}, {Dobler}, {Donalek}, {Dubois-Felsmann}, {Durech},
  {Eliasdottir}, {Eracleous}, {Eyer}, {Falco}, {Fan}, {Fassnacht}, {Ferguson},
  {Fernandez}, {Fields}, {Finkbeiner}, {Figueroa}, {Fox}, {Francke}, {Frank},
  {Frieman}, {Fromenteau}, {Furqan}, {Galaz}, {Gal-Yam}, {Garnavich},
  {Gawiser}, {Geary}, {Gee}, {Gibson}, {Gilmore}, {Grace}, {Green}, {Gressler},
  {Grillmair}, {Habib}, {Haggerty}, {Hamuy}, {Harris}, {Hawley}, {Heavens},
  {Hebb}, {Henry}, {Hileman}, {Hilton}, {Hoadley}, {Holberg}, {Holman},
  {Howell}, {Infante}, {Ivezic}, {Jacoby}, {Jain}, {R}, {Jedicke}, {Jee},
  {Garrett Jernigan}, {Jha}, {Johnston}, {Jones}, {Juric}, {Kaasalainen},
  {Styliani}, {Kafka}, {Kahn}, {Kaib}, {Kalirai}, {Kantor}, {Kasliwal},
  {Keeton}, {Kessler}, {Knezevic}, {Kowalski}, {Krabbendam}, {Krughoff},
  {Kulkarni}, {Kuhlman}, {Lacy}, {Lepine}, {Liang}, {Lien}, {Lira}, {Long},
  {Lorenz}, {Lotz}, {Lupton}, {Lutz}, {Macri}, {Mahabal}, {Mandelbaum},
  {Marshall}, {May}, {McGehee}, {Meadows}, {Meert}, {Milani}, {Miller},
  {Miller}, {Mills}, {Minniti}, {Monet}, {Mukadam}, {Nakar}, {Neill}, {Newman},
  {Nikolaev}, {Nordby}, {O'Connor}, {Oguri}, {Oliver}, {Olivier}, {Olsen},
  {Olsen}, {Olszewski}, {Oluseyi}, {Padilla}, {Parker}, {Pepper}, {Peterson},
  {Petry}, {Pinto}, {Pizagno}, {Popescu}, {Prsa}, {Radcka}, {Raddick},
  {Rasmussen}, {Rau}, {Rho}, {Rhoads}, {Richards}, {Ridgway}, {Robertson},
  {Roskar}, {Saha}, {Sarajedini}, {Scannapieco}, {Schalk}, {Schindler}, \&
  {Schmidt}}]{LSSTScienceBook2009arXiv0912.0201L}
{LSST Science Collaboration}, {Abell}, P.~A., {Allison}, J., {et~al.} 2009,
  arXiv e-prints, arXiv:0912.0201, \dodoi{10.48550/arXiv.0912.0201}

\bibitem[{{MacLeod} {et~al.}(2019){MacLeod}, {Green}, {Anderson}, {Bruce},
  {Eracleous}, {Graham}, {Homan}, {Lawrence}, {LeBleu}, {Ross}, {Ruan},
  {Runnoe}, {Stern}, {Burgett}, {Chambers}, {Kaiser}, {Magnier}, \&
  {Metcalfe}}]{MacLeod_2019ApJ...874....8M}
{MacLeod}, C.~L., {Green}, P.~J., {Anderson}, S.~F., {et~al.} 2019, \apj, 874,
  8, \dodoi{10.3847/1538-4357/ab05e2}

\bibitem[{{Magdziarz} {et~al.}(1998){Magdziarz}, {Blaes}, {Zdziarski},
  {Johnson}, \& {Smith}}]{Magdziarz1998MNRAS.301..179M}
{Magdziarz}, P., {Blaes}, O.~M., {Zdziarski}, A.~A., {Johnson}, W.~N., \&
  {Smith}, D.~A. 1998, \mnras, 301, 179,
  \dodoi{10.1046/j.1365-8711.1998.02015.x}

\bibitem[{{Mainieri} {et~al.}(2024){Mainieri}, {Anderson}, {Brinchmann},
  {Cimatti}, {Ellis}, {Hill}, {Kneib}, {McLeod}, {Opitom}, {Roth},
  {Sanchez-Saez}, {Smiljanic}, {Tolstoy}, {Bacon}, {Randich}, {Adamo},
  {Annibali}, {Arevalo}, {Audard}, {Barsanti}, {Battaglia}, {Bayo Aran},
  {Belfiore}, {Bellazzini}, {Bellini}, {Beltran}, {Berni}, {Bianchi}, {Biazzo},
  {Bisero}, {Bisogni}, {Bland-Hawthorn}, {Blondin}, {Bodensteiner}, {Boffin},
  {Bonito}, {Bono}, {Bouche}, {Bowman}, {Braga}, {Bragaglia}, {Branchesi},
  {Brucalassi}, {Bryant}, {Bryson}, {Busa}, {Camera}, {Carbone}, {Casali},
  {Casali}, {Casasola}, {Castro}, {Catelan}, {Cavallo}, {Chiappini}, {Cioni},
  {Colless}, {Colzi}, {Contarini}, {Couch}, {D'Ammando}, {d'Assignies D.},
  {D'Orazi}, {da Silva}, {Dainotti}, {Damiani}, {Danielski}, {De Cia}, {de
  Jong}, {Dhawan}, {Dierickx}, {Driver}, {Dupletsa}, {Escoffier}, {Escorza},
  {Fabrizio}, {Fiorentino}, {Fontana}, {Fontani}, {Forero Sanchez}, {Franois},
  {Galindo-Guil}, {Gallazzi}, {Galli}, {Garcia}, {Garcia-Rojas}, {Garilli},
  {Grand}, {Guarcello}, {Hazra}, {Helmi}, {Herrero}, {Iglesias}, {Ilic},
  {Irsic}, {Ivanov}, {Izzo}, {Jablonka}, {Joachimi}, {Kakkad}, {Kamann},
  {Koposov}, {Kordopatis}, {Kovacevic}, {Kraljic}, {Kuncarayakti}, {Kwon}, {La
  Forgia}, {Lahav}, {Laigle}, {Lazzarin}, {Leaman}, {Leclercq}, {Lee}, {Lee},
  {Lehnert}, {Lira}, {Loffredo}, {Lucatello}, {Magrini}, {Maguire}, {Mahler},
  {Zahra Majidi}, {Malavasi}, {Mannucci}, {Marconi}, {Martin}, {Marulli},
  {Massari}, {Matsuno}, {Mattheee}, {McGee}, {Merc}, {Merle}, {Miglio},
  {Migliorini}, {Minchev}, {Minniti}, {Miret-Roig}, {Monreal Ibero}, {Montano},
  {Montet}, {Moresco}, {Moretti}, {Moscardini}, {Moya}, {Mueller},
  {Nanayakkara}, {Nicholl}, {Nordlander}, {Onori}, {Padovani}, {Pala}, {Panda},
  {Pandey-Pommier}, {Pasquini}, {Pawlak}, {Pessi}, {Pisani}, {Popovic},
  {Prisinzano}, {Raddi}, {Rainer}, {Rebassa-Mansergas}, {Richard}, {Rigault},
  {Rocher}, {Romano}, {Rosati}, {Sacco}, {Sanchez-Janssen}, {Sander},
  {Sanders}, {Sargent}, {Sarpa}, {Schimd}, {Schipani}, {Sefusatti}, {Smith},
  {Spina}, {Steinmetz}, {Tacchella}, {Tautvaisiene}, {Theissen}, {Thomas},
  {Ting}, {Travouillon}, {Tresse}, {Trivedi}, {Tsantaki}, {Tsedrik}, {Urrutia},
  {Valenti}, {Van der Swaelmen}, {Van Eck}, {Verdiani}, {Verdier}, {Vergani},
  {Verhamme}, \& {Vernet}}]{WST_2024arXiv240305398M}
{Mainieri}, V., {Anderson}, R.~I., {Brinchmann}, J., {et~al.} 2024, arXiv
  e-prints, arXiv:2403.05398, \dodoi{10.48550/arXiv.2403.05398}

\bibitem[{{Mandal} {et~al.}(2025){Mandal}, {Pozo Nu{\~n}ez}, {Jaiswal},
  {Naddaf}, {Czerny}, {Panda}, {Karczmarek}, {Pietrzy{\'n}ski}, {Pandey},
  {Peterson}, {Zaja{\v{c}}ek}, {Dov{\v{c}}iak}, {Karas}, {Narloch}, {Kicia},
  {G{\'o}rski}, {Ka{\l}uszy{\'n}ski}, {Hajdu}, {Wielg{\'o}rski}, {Zgirski},
  {Ga{\l}an}, {Pych}, {Smolec}, {B{\k{a}}kowska}, {Gieren}, \&
  {Kervella}}]{Mandal2025arXiv251213296M}
{Mandal}, A.~K., {Pozo Nu{\~n}ez}, F., {Jaiswal}, V.~K., {et~al.} 2025, arXiv
  e-prints, arXiv:2512.13296, \dodoi{10.48550/arXiv.2512.13296}

\bibitem[{{Marziani} {et~al.}(2025){Marziani}, {Bon}, {Bon}, \&
  {D'Onofrio}}]{Marziani2025Univ...11...76M}
{Marziani}, P., {Bon}, E., {Bon}, N., \& {D'Onofrio}, M. 2025, Universe, 11,
  76, \dodoi{10.3390/universe11030076}

\bibitem[{{Mezcua}(2017)}]{Mezcua2017IJMPD..2630021M}
{Mezcua}, M. 2017, International Journal of Modern Physics D, 26, 1730021,
  \dodoi{10.1142/S021827181730021X}

\bibitem[{{Netzer}(2022)}]{Netzer2022MNRAS.509.2637N}
{Netzer}, H. 2022, \mnras, 509, 2637, \dodoi{10.1093/mnras/stab3133}

\bibitem[{{Neustadt} {et~al.}(2024){Neustadt}, {Kochanek}, {Montano},
  {Gelbord}, {Barth}, {De Rosa}, {Kriss}, {Cackett}, {Horne}, {Kara}, {Landt},
  {Netzer}, {Arav}, {Bentz}, {Dalla Bont{\`a}}, {Dehghanian}, {Du}, {Edelson},
  {Ferland}, {Fian}, {Fischer}, {Goad}, {Gonz{\'a}lez Buitrago}, {Gorjian},
  {Grier}, {Hall}, {Homayouni}, {Hu}, {Ili{\'c}}, {Joner}, {Kaastra}, {Kaspi},
  {Korista}, {Kova{\v{c}}evi{\'c}}, {Lewin}, {Li}, {McHardy}, {Mehdipour},
  {Miller}, {Panagiotou}, {Partington}, {Plesha}, {Pogge}, {Popovi{\'c}},
  {Proga}, {Storchi-Bergmann}, {Sanmartim}, {Siebert}, {Signorini},
  {Vestergaard}, {Zaidouni}, \& {Zu}}]{Neustadt_2024ApJ...961..219N}
{Neustadt}, J. M.~M., {Kochanek}, C.~S., {Montano}, J., {et~al.} 2024, \apj,
  961, 219, \dodoi{10.3847/1538-4357/ad1386}

\bibitem[{{Noda} \& {Done}(2018)}]{Noda_2018MNRAS.480.3898N}
{Noda}, H., \& {Done}, C. 2018, \mnras, 480, 3898,
  \dodoi{10.1093/mnras/sty2032}

\bibitem[{{Panda} {et~al.}(2019){Panda}, {Mart{\'\i}nez-Aldama}, \&
  {Zaja{\v{c}}ek}}]{Panda_2019FrASS...6...75P}
{Panda}, S., {Mart{\'\i}nez-Aldama}, M.~L., \& {Zaja{\v{c}}ek}, M. 2019,
  Frontiers in Astronomy and Space Sciences, 6, 75,
  \dodoi{10.3389/fspas.2019.00075}

\bibitem[{{Panda} {et~al.}(2024){Panda}, {Pozo Nu{\~n}ez}, {Ba{\~n}ados}, \&
  {Heidt}}]{Panda_2024ApJ...968L..16P}
{Panda}, S., {Pozo Nu{\~n}ez}, F., {Ba{\~n}ados}, E., \& {Heidt}, J. 2024,
  \apjl, 968, L16, \dodoi{10.3847/2041-8213/ad5014}

\bibitem[{{Panda} \& {{\'S}niegowska}(2024)}]{Panda_2024ApJS..272...13P}
{Panda}, S., \& {{\'S}niegowska}, M. 2024, \apjs, 272, 13,
  \dodoi{10.3847/1538-4365/ad344f}

\bibitem[{{Panda} {et~al.}(2025){Panda}, {Benati Gon{\c{c}}alves},
  {Storchi-Bergmann}, {{\'S}niegowska}, {Czerny}, {Bon}, {Marziani}, {Bon},
  {Rodr{\'\i}guez Ardila}, {May}, {Fonseca Far{\'\i}a}, {Fraga}, {Pozo
  Nu{\~n}ez}, {Ba{\~n}ados}, {Heidt}, {Garnica}, \&
  {Dultzin}}]{Panda2025arXiv251001486P}
{Panda}, S., {Benati Gon{\c{c}}alves}, H., {Storchi-Bergmann}, T., {et~al.}
  2025, arXiv e-prints, arXiv:2510.01486, \dodoi{10.48550/arXiv.2510.01486}

\bibitem[{{Peca} {et~al.}(2024){Peca}, {Cappelluti}, {LaMassa}, {Urry},
  {Moscetti}, {Marchesi}, {Sanders}, {Auge}, {Ghosh}, {Ananna},
  {Torres-Alb{\`a}}, \& {Treister}}]{Peca2024ApJ...974..156P}
{Peca}, A., {Cappelluti}, N., {LaMassa}, S., {et~al.} 2024, \apj, 974, 156,
  \dodoi{10.3847/1538-4357/ad6df4}

\bibitem[{{Pozo Nu{\~n}ez} {et~al.}(2025){Pozo Nu{\~n}ez}, {Ba{\~n}ados},
  {Panda}, \& {Heidt}}]{FPN_2025A&A...700L...8P}
{Pozo Nu{\~n}ez}, F., {Ba{\~n}ados}, E., {Panda}, S., \& {Heidt}, J. 2025,
  \aap, 700, L8, \dodoi{10.1051/0004-6361/202555421}

\bibitem[{{Pozo Nu{\~n}ez} {et~al.}(2023){Pozo Nu{\~n}ez}, {Bruckmann},
  {Deesamutara}, {Czerny}, {Panda}, {Lobban}, {Pietrzy{\'n}ski}, \&
  {Polsterer}}]{FPN2023MNRAS.522.2002P}
{Pozo Nu{\~n}ez}, F., {Bruckmann}, C., {Deesamutara}, S., {et~al.} 2023,
  \mnras, 522, 2002, \dodoi{10.1093/mnras/stad286}

\bibitem[{{Pozo Nu{\~n}ez} {et~al.}(2024){Pozo Nu{\~n}ez}, {Czerny}, {Panda},
  {Kovacevic}, {Brandt}, {Horne}, \& {LSST AGN Science
  Collaboration}}]{FPN_2024RNAAS...8...47P}
{Pozo Nu{\~n}ez}, F., {Czerny}, B., {Panda}, S., {et~al.} 2024, Research Notes
  of the American Astronomical Society, 8, 47, \dodoi{10.3847/2515-5172/ad284a}

\bibitem[{{Pozo Nu{\~n}ez} {et~al.}(2014){Pozo Nu{\~n}ez}, {Haas}, {Ramolla},
  {Bruckmann}, {Westhues}, {Chini}, {Steenbrugge}, {Lemke}, {Murphy}, \&
  {Kollatschny}}]{FPN2014A&A...568A..36P}
{Pozo Nu{\~n}ez}, F., {Haas}, M., {Ramolla}, M., {et~al.} 2014, \aap, 568, A36,
  \dodoi{10.1051/0004-6361/201322736}

\bibitem[{{Prince} {et~al.}(2023){Prince}, {Zaja{\v{c}}ek}, {Panda},
  {Hryniewicz}, {Kumar Jaiswal}, {Czerny}, {Trzcionkowski}, {Bronikowski},
  {Ra{\l}owski}, {Sobrino Figaredo}, {Martinez-Aldama}, {{\'S}niegowska},
  {{\'S}redzi{\'n}ska}, {Bilicki}, {Naddaf}, {Pandey}, {Haas}, {Sarna},
  {Pietrzy{\'n}ski}, {Karas}, {Olejak}, {Przy{\l}uski}, {Sefako}, {Genade},
  {Worters}, {Koz{\l}owski}, \& {Udalski}}]{Prince2023A&A...678A.189P}
{Prince}, R., {Zaja{\v{c}}ek}, M., {Panda}, S., {et~al.} 2023, \aap, 678, A189,
  \dodoi{10.1051/0004-6361/202346738}

\bibitem[{{Rees}(1988)}]{Rees1988Natur.333..523R}
{Rees}, M.~J. 1988, \nat, 333, 523, \dodoi{10.1038/333523a0}

\bibitem[{{Ren} {et~al.}(2022){Ren}, {Wang}, {Cai}, \&
  {Guo}}]{Ren_2022ApJ...925...50R}
{Ren}, W., {Wang}, J., {Cai}, Z., \& {Guo}, H. 2022, \apj, 925, 50,
  \dodoi{10.3847/1538-4357/ac3828}

\bibitem[{{Ricci} \& {Trakhtenbrot}(2023)}]{Ricci_2023NatAs...7.1282R}
{Ricci}, C., \& {Trakhtenbrot}, B. 2023, Nature Astronomy, 7, 1282,
  \dodoi{10.1038/s41550-023-02108-4}

\bibitem[{{Runnoe} {et~al.}(2016){Runnoe}, {Cales}, {Ruan}, {Eracleous},
  {Anderson}, {Shen}, {Green}, {Morganson}, {LaMassa}, {Greene}, {Dwelly},
  {Schneider}, {Merloni}, {Georgakakis}, \&
  {Roman-Lopes}}]{Runnoe_2016MNRAS.455.1691R}
{Runnoe}, J.~C., {Cales}, S., {Ruan}, J.~J., {et~al.} 2016, \mnras, 455, 1691,
  \dodoi{10.1093/mnras/stv2385}

\bibitem[{{S{\'a}nchez-S{\'a}ez} {et~al.}(2021){S{\'a}nchez-S{\'a}ez}, {Lira},
  {Mart{\'\i}}, {S{\'a}nchez-Pi}, {Arredondo}, {Bauer}, {Bayo},
  {Cabrera-Vives}, {Donoso-Oliva}, {Est{\'e}vez}, {Eyheramendy}, {F{\"o}rster},
  {Hern{\'a}ndez-Garc{\'\i}a}, {Arancibia}, {P{\'e}rez-Carrasco},
  {Sep{\'u}lveda}, \& {Vergara}}]{SanchezSaez2021AJ....162..206S}
{S{\'a}nchez-S{\'a}ez}, P., {Lira}, H., {Mart{\'\i}}, L., {et~al.} 2021, \aj,
  162, 206, \dodoi{10.3847/1538-3881/ac1426}

\bibitem[{{S{\'a}nchez-S{\'a}ez} {et~al.}(2024){S{\'a}nchez-S{\'a}ez},
  {Hern{\'a}ndez-Garc{\'\i}a}, {Bernal}, {Bayo}, {Calistro Rivera}, {Bauer},
  {Ricci}, {Merloni}, {Graham}, {Cartier}, {Ar{\'e}valo}, {Assef}, {Concas},
  {Homan}, {Krumpe}, {Lira}, {Malyali}, {Mart{\'\i}nez-Aldama}, {Mu{\~n}oz
  Arancibia}, {Rau}, {Bruni}, {F{\"o}rster}, {Pavez-Herrera},
  {Tub{\'\i}n-Arenas}, \& {Brightman}}]{Sanchez-Saez2024A&A...688A.157S}
{S{\'a}nchez-S{\'a}ez}, P., {Hern{\'a}ndez-Garc{\'\i}a}, L., {Bernal}, S.,
  {et~al.} 2024, \aap, 688, A157, \dodoi{10.1051/0004-6361/202347957}

\bibitem[{{Shakura} \& {Sunyaev}(1973)}]{SS_1973A&A....24..337S}
{Shakura}, N.~I., \& {Sunyaev}, R.~A. 1973, \aap, 24, 337

\bibitem[{{Shankar} {et~al.}(2009){Shankar}, {Weinberg}, \&
  {Miralda-Escud{\'e}}}]{Shankar2009ApJ...690...20S}
{Shankar}, F., {Weinberg}, D.~H., \& {Miralda-Escud{\'e}}, J. 2009, \apj, 690,
  20, \dodoi{10.1088/0004-637X/690/1/20}

\bibitem[{{Shen} {et~al.}(2020){Shen}, {Hopkins}, {Faucher-Gigu{\`e}re},
  {Alexander}, {Richards}, {Ross}, \& {Hickox}}]{Shen2020MNRAS.495.3252S}
{Shen}, X., {Hopkins}, P.~F., {Faucher-Gigu{\`e}re}, C.-A., {et~al.} 2020,
  \mnras, 495, 3252, \dodoi{10.1093/mnras/staa1381}

\bibitem[{{Shen} {et~al.}(2024){Shen}, {Grier}, {Horne}, {Stone}, {Li}, {Yang},
  {Homayouni}, {Trump}, {Anderson}, {Brandt}, {Hall}, {Ho}, {Jiang},
  {Petitjean}, {Schneider}, {Tao}, {Donnan}, {AlSayyad}, {Bershady}, {Blanton},
  {Bizyaev}, {Bundy}, {Chen}, {Davis}, {Dawson}, {Fan}, {Greene},
  {Gr{\"o}ller}, {Guo}, {Ibarra-Medel}, {Jiang}, {Keenan}, {Kollmeier},
  {Lejoly}, {Li}, {de la Macorra}, {Moe}, {Nie}, {Rossi}, {Smith}, {Tee},
  {Weijmans}, {Xu}, {Yue}, {Zhou}, {Zhou}, \& {Zou}}]{Shen2024ApJS..272...26S}
{Shen}, Y., {Grier}, C.~J., {Horne}, K., {et~al.} 2024, \apjs, 272, 26,
  \dodoi{10.3847/1538-4365/ad3936}

\bibitem[{{Sniegowska} {et~al.}(2020){Sniegowska}, {Czerny}, {Bon}, \&
  {Bon}}]{Sniegowska_2020A&A...641A.167S}
{Sniegowska}, M., {Czerny}, B., {Bon}, E., \& {Bon}, N. 2020, \aap, 641, A167,
  \dodoi{10.1051/0004-6361/202038575}

\bibitem[{{Stern} {et~al.}(2018){Stern}, {McKernan}, {Graham}, {Ford}, {Ross},
  {Meisner}, {Assef}, {Balokovi{\'c}}, {Brightman}, {Dey}, {Drake},
  {Djorgovski}, {Eisenhardt}, \& {Jun}}]{Stern_2018ApJ...864...27S}
{Stern}, D., {McKernan}, B., {Graham}, M.~J., {et~al.} 2018, \apj, 864, 27,
  \dodoi{10.3847/1538-4357/aac726}

\bibitem[{{Su} {et~al.}(2025){Su}, {Cai}, {Guo}, {Sun}, \&
  {Wang}}]{Su2025ApJ...990...10S}
{Su}, Z.-B., {Cai}, Z.-Y., {Guo}, H., {Sun}, M., \& {Wang}, J.-X. 2025, \apj,
  990, 10, \dodoi{10.3847/1538-4357/adef13}

\bibitem[{{Suganuma} {et~al.}(2006){Suganuma}, {Yoshii}, {Kobayashi},
  {Minezaki}, {Enya}, {Tomita}, {Aoki}, {Koshida}, \&
  {Peterson}}]{Sugunuma_2006ApJ...639...46S}
{Suganuma}, M., {Yoshii}, Y., {Kobayashi}, Y., {et~al.} 2006, \apj, 639, 46,
  \dodoi{10.1086/499326}

\bibitem[{{Temple} {et~al.}(2023){Temple}, {Ricci}, {Koss}, {Trakhtenbrot},
  {Bauer}, {Mushotzky}, {Rojas}, {Caglar}, {Harrison}, {Oh}, {Padilla
  Gonzalez}, {Powell}, {Ricci}, {Riffel}, {Stern}, \&
  {Urry}}]{Temple_2023MNRAS.518.2938T}
{Temple}, M.~J., {Ricci}, C., {Koss}, M.~J., {et~al.} 2023, \mnras, 518, 2938,
  \dodoi{10.1093/mnras/stac3279}

\bibitem[{{Trakhtenbrot} {et~al.}(2019){Trakhtenbrot}, {Arcavi}, {Ricci},
  {Tacchella}, {Stern}, {Netzer}, {Jonker}, {Horesh}, {Mej{\'\i}a-Restrepo},
  {Hosseinzadeh}, {Hallefors}, {Howell}, {McCully}, {Balokovi{\'c}}, {Heida},
  {Kamraj}, {Lansbury}, {Wyrzykowski}, {Gromadzki}, {Hamanowicz}, {Cenko},
  {Sand}, {Hsiao}, {Phillips}, {Diamond}, {Kara}, {Gendreau}, {Arzoumanian}, \&
  {Remillard}}]{Trakhtenbrot2019NatAs...3..242T}
{Trakhtenbrot}, B., {Arcavi}, I., {Ricci}, C., {et~al.} 2019, Nature Astronomy,
  3, 242, \dodoi{10.1038/s41550-018-0661-3}

\bibitem[{{van Velzen} {et~al.}(2020){van Velzen}, {Holoien}, {Onori}, {Hung},
  \& {Arcavi}}]{vanVelzen2020SSRv..216..124V}
{van Velzen}, S., {Holoien}, T. W.-S., {Onori}, F., {Hung}, T., \& {Arcavi}, I.
  2020, \ssr, 216, 124, \dodoi{10.1007/s11214-020-00753-z}

\bibitem[{{Wang} {et~al.}(2023){Wang}, {Guo}, \&
  {Woo}}]{ShuWang2023ApJ...948L..23W}
{Wang}, S., {Guo}, H., \& {Woo}, J.-H. 2023, \apjl, 948, L23,
  \dodoi{10.3847/2041-8213/accf96}

\bibitem[{{Wang} {et~al.}(2024){Wang}, {Woo}, {Gallo}, {Guo}, {Son}, {Kong},
  {Mandal}, {Cho}, {Kim}, \& {Shin}}]{ShuWang_2024ApJ...966..128W}
{Wang}, S., {Woo}, J.-H., {Gallo}, E., {et~al.} 2024, \apj, 966, 128,
  \dodoi{10.3847/1538-4357/ad3049}

\bibitem[{{Wiseman} {et~al.}(2025){Wiseman}, {Williams}, {Arcavi}, {Galbany},
  {Graham}, {H{\"o}nig}, {Newsome}, {Subrayan}, {Sullivan}, {Wang}, {Ili{\'c}},
  {Nicholl}, {Oates}, {Petrushevska}, \& {Smith}}]{Wiseman2025MNRAS.537.2024W}
{Wiseman}, P., {Williams}, R.~D., {Arcavi}, I., {et~al.} 2025, \mnras, 537,
  2024, \dodoi{10.1093/mnras/staf116}

\bibitem[{{Yu} {et~al.}(2025){Yu}, {Ruan}, {Burke}, {Assef}, {Ananna}, {Bauer},
  {De Cicco}, {Horne}, {Hern{\'a}ndez-Garc{\'\i}a}, {Ili{\'c}}, {Jha},
  {Kova{\v{c}}evi{\'c}}, {Marculewicz}, {Panda}, {Ricci}, {Richards}, {Riffel},
  {Schneider}, {S{\'a}nchez-S{\'a}ez}, {Satheesh Sheeba}, {Tombesi}, {Temple},
  {Vogeley}, {Yoon}, \& {Zou}}]{Yu2025arXiv251121479Y}
{Yu}, W., {Ruan}, J.~J., {Burke}, C.~J., {et~al.} 2025, arXiv e-prints,
  arXiv:2511.21479, \dodoi{10.48550/arXiv.2511.21479}

\bibitem[{{Zeltyn} {et~al.}(2024){Zeltyn}, {Trakhtenbrot}, {Eracleous}, {Yang},
  {Green}, {Anderson}, {LaMassa}, {Runnoe}, {Assef}, {Bauer}, {Brandt},
  {Davis}, {Frederick}, {Fries}, {Graham}, {Grogin}, {Guolo},
  {Hern{\'a}ndez-Garc{\'\i}a}, {Koekemoer}, {Krumpe}, {Liu},
  {Mart{\'\i}nez-Aldama}, {Ricci}, {Schneider}, {Shen}, {{\'S}niegowska},
  {Temple}, {Trump}, {Xue}, {Brownstein}, {Dwelly}, {Morrison}, {Bizyaev},
  {Pan}, \& {Kollmeier}}]{Zeltyn_2024ApJ...966...85Z}
{Zeltyn}, G., {Trakhtenbrot}, B., {Eracleous}, M., {et~al.} 2024, \apj, 966,
  85, \dodoi{10.3847/1538-4357/ad2f30}

\bibitem[{{Zou} {et~al.}(2022){Zou}, {Brandt}, {Chen}, {Leja}, {Ni}, {Yan},
  {Yang}, {Zhu}, {Luo}, {Nyland}, {Vito}, \& {Xue}}]{Zou2022ApJS..262...15Z}
{Zou}, F., {Brandt}, W.~N., {Chen}, C.-T., {et~al.} 2022, \apjs, 262, 15,
  \dodoi{10.3847/1538-4365/ac7bdf}

\end{thebibliography}

\section{ACKNOWLEDGEMENTS}
We thank Roberto Assef (UDP, Chile) for helpful discussions on the quasar number counts and notes on the latest cadence simulations for the Rubin Observatory. We are also grateful to Eduardo Ba\~nados (MPIA, Germany) and the members of the LSST AGN Science Collaboration for fruitful discussions. SP is supported by the international Gemini Observatory, a program of NSF NOIRLab, which is managed by the Association of Universities for Research in Astronomy (AURA) under a cooperative agreement with the U.S. National Science Foundation, on behalf of the Gemini partnership of Argentina, Brazil, Canada, Chile, the Republic of Korea, and the United States of America.

\section{APPENDIX}

\begin{figure}[!htb]
    \centering
    \includegraphics[width=\linewidth, trim={0 0 0 1.95cm},clip]{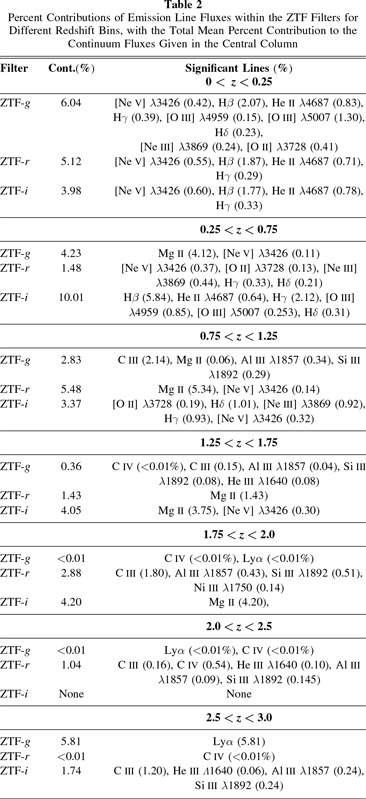}
    \caption{\justifying Percent emission-line flux contributions within the ZTF filters across redshift bins; the central column lists the total mean percent contribution to the continuum flux. Courtesy: \citet{Hygor_2025ApJ...988...27B}.}
    \label{fig:ztf_cont_alt}
\end{figure}



\end{document}